\documentclass[a4paper,fleqn]{cas-dc}
\usepackage[authoryear]{natbib}
\usepackage[figuresright]{rotating}
\usepackage{moreverb}
\usepackage{amsmath,amssymb,amsfonts}
\usepackage{algorithmic}
\usepackage{url}

\usepackage{graphicx}
\usepackage{textcomp}
\usepackage{xcolor}
\usepackage{tcolorbox}

\usepackage{ragged2e}
\usepackage{balance}
\usepackage{colortbl}
\usepackage{multirow}
\usepackage{color}

\usepackage{CJKutf8}
\usepackage{siunitx}
\usepackage{rotfloat}
\usepackage{float}
\usepackage{lscape}
\usepackage{amssymb,mathrsfs,amsmath}
\usepackage{adjustbox}
\usepackage{array} 
\def\tsc#1{\csdef{#1}{\textsc{\lowercase{#1}}\xspace}}
\tsc{WGM}
\tsc{QE}
\tsc{EP}
\tsc{PMS}
\tsc{BEC}
\tsc{DE}
\begin{document}
\begin{sloppypar}
% \def\UrlOrds{\do\\-}
% \g@addto@macro{\UrlBreaks}{\UrlOrds}

\let\WriteBookmarks\relax
\def\floatpagepagefraction{1}
\def\textpagefraction{.001}
\shorttitle{Problems, Causes and Solutions of Using GitHub Copilot}
\shortauthors{X Zhou et al.}
\title [mode = title]{Exploring the Problems, their Causes and Solutions of AI Pair Programming: A Study on GitHub and Stack Overflow}

\author[1]{Xiyu Zhou}
\ead{xiyuzhou@whu.edu.cn}
\credit{Conceptualization, Investigation, Data curation, Formal analysis, Writing - Original draft preparation}

\author[1]{Peng Liang}
\cormark[1]
\ead{liangp@whu.edu.cn}
\credit{Conceptualization, Methodology, Investigation, Data curation, Supervision, Writing - review and editing}
\address[1]{School of Computer Science, Wuhan University, Wuhan, China}

\author[1]{Beiqi Zhang}
\ead{zhangbeiqi@whu.edu.cn}
\credit{Investigation, Data curation, Formal Analysis, Writing - Original draft preparation}

\author[2]{Zengyang Li}
\ead{zengyangli@ccnu.edu.cn}
\credit{Conceptualization, Methodology, Writing - review and editing}
\address[2]{School of Computer Science, Central China Normal University, Wuhan, China}

\author[3]{Aakash Ahmad}
\ead{ahmad.aakash@gmail.com}
\credit{Conceptualization, Methodology, Writing - review and editing}
\address[3]{School of Computing and Communications, Lancaster University Leipzig, Leipzig, Germany}

\author[4]{Mojtaba Shahin}
\ead{mojtaba.shahin@rmit.edu.au}
\credit{Conceptualization, Methodology, Writing - review and editing}
\address[4]{School of Computing Technologies, RMIT University, Melbourne, Australia}

\author[5]{Muhammad Waseem}
\ead{muhammad.m.waseem@jyu.fi}
\credit{Methodology, Writing - review and editing}
\address[5]{Faculty of Information Technology, University of Jyväskylä, Jyväskylä, Finland}

\cortext[cor1]{Corresponding author.}

%\fundinginfo{Natural Science Foundation of Hubei
%Province of China, Grant Number: 2021CFB577\\
%National Natural Science Foundation of China, Grant Number: 62176099}

\begin{abstract}
With the recent advancement of Artificial Intelligence (AI) and Large Language Models (LLMs), AI-based code generation tools become a practical solution for software development. GitHub Copilot, the AI pair programmer, utilizes machine learning models trained on a large corpus of code snippets to generate code suggestions using natural language processing.
%\textit{Aims}: 
Despite its popularity in software development, there is limited empirical evidence on the actual experiences of practitioners who work with Copilot. 
%\textit{Method}: 
To this end, we conducted an empirical study to understand the problems that practitioners face when using Copilot, as well as their underlying causes and potential solutions. We collected data from 473 GitHub issues, 706 GitHub discussions, and 142 Stack Overflow posts.
%\textit{Results}: 
Our results reveal that (1) \textit{Operation Issue} and \textit{Compatibility Issue} are the most common problems faced by Copilot users, (2) \textit{Copilot Internal Error}, \textit{Network Connection Error}, and \textit{Editor/IDE Compatibility Issue} are identified as the most frequent causes, and (3) \textit{Bug Fixed by Copilot}, \textit{Modify Configuration/Setting}, and \textit{Use Suitable Version} are the predominant solutions.
Based on the results, we discuss the potential areas of Copilot for enhancement, and provide the implications for the Copilot users, the Copilot team, and researchers.
\end{abstract}

\begin{keywords}
GitHub Copilot, GitHub Issues, GitHub Discussions, StackOverflow Post, Problem, Cause, Solution
\end{keywords}

\maketitle

\section{Introduction}
\label{sec:introduction}
In software development, developers strive to achieve automation and intelligence to generate most code automatically with minimal human coding effort. Several studies (e.g., \cite{sifei2019aroma}, \cite{martin2010recommendation}) and software products (e.g., Eclipse Code Recommenders \citep{eclipsecoderecommenders}) have been dedicated to improving the efficiency of developers through the development of systems that can recommend and generate code. However, early AI code generation tools mostly used heuristic rules or expert systems, which faced significant limitations, including inflexibility and challenges in scalability \citep{jiang2024asurvey}, making them less adaptable to varying contexts. Large Language Models (LLMs) are a type of natural language processing (NLP) technique based on deep learning that is capable of automatically learning the grammar, semantics, and pragmatics of language, and generating a wide variety of contents. Due to the extensive number of parameters and large-scale training datasets, LLMs have demonstrated powerful capabilities in NLP, often approaching or even surpassing human-level performance in NLP tasks such as text translation \citep{hui2023towards} and sentiment analysis \citep{amin2023will} \citep{chen2023robust}. The expansion of model parameters also enables LLMs to capture intricate language patterns, resulting in sustained performance enhancements across a wide range of downstream tasks \citep{wei2022emergent}. Therefore, the emergence of LLMs has significantly reshaped the field of code generation tasks \citep{chen2021evaluating}. Recently, AI code generation tools driven by LLMs that have been trained on large amounts of code snippets are increasingly in the spotlight (e.g., AI-augmented development in Gartner Trends 2024 \citep{gartner2024}), making it possible for programmers to automatically generate code with minimized human effort \citep{austin2021program}.

On June 29, 2021, GitHub and OpenAI jointly announced the launch of a new product named GitHub Copilot \citep{githubcopilot}. This innovative tool is powered by OpenAI's Codex, a large-scale neural network model that is trained on a massive dataset of source code and natural language text. The goal of GitHub Copilot is to provide advanced code autocompletion and generation capabilities to developers, effectively acting as an ``AI pair programmer'' that can assist with coding tasks in real-time. Copilot has been designed to work with a wide range of Integrated Development Environments (IDEs) and code editors, such as VSCode, Visual Studio, Neovim, and JetBrains \citep{githubcopilot}. By collecting contextual information like function names and comments, Copilot is able to generate code snippets in a variety of programming languages (e.g., Python, C++, Java), which can improve developers' productivity and help them complete coding tasks more efficiently \citep{imai2022github}.

Since its release, Copilot has gained significant attention within the developer community, and it had a total of 1.3 million paid users till Feb of 2024~\citep{copilotusers}. Many studies identify the effectiveness and concerns about the potential impact on code security and intellectual property \citep{hammond2022asleep} \citep{bird2022taking} \citep{jaworski2023study}. Some prior research investigated the quality of the code generated by Copilot \citep{yetistiren2022assessing} \citep{nguyen2022evalution}, while others examined its performance in practical software development \citep{imai2022github} \citep{shraddha2023grounded} \citep{peng2023impact}. 

%However, little is known about the problems, causes, and solutions encountered during the practical use of Copilot. 
However, there is currently a lack of systematic categorization of the problems arise during the practical use of Copilot from the perspective of developers, as well as the causes behind them and solutions for addressing them.
%Considering that GitHub Copilot is a widely used and highly representative AI code generation tool, it is important to investigate the various challenges users encounter, as well as their causes and solutions. This study evaluates the usefulness of Copilot in different aspects, as well as explore the ways in which such AI code generation tools interact with software developers. 
To this end, we conducted a thorough analysis of the problems faced by software developers when coding with GitHub Copilot, as well as their causes and solutions, by collecting data from GitHub Issues, GitHub Discussions, and Stack Overflow (SO) posts, which would help to understand the limitations of Copilot in practical settings. 
%To be more precise, we formulated three Research Questions (RQs) to direct our study: 
%\textbf{RQ1: What are the issues faced by users while using Copilot in practice software development?} 
%\textbf{RQ2: What are the underlying causes of these issues?} 
%\textbf{RQ3: What are the potential solutions to address these issues?}

%\textcolor{blue}{After specifying the research questions, We retrieved the issues, discussions, and posts potentially related to GitHub Copilot through official APIs. Then we conducted data labelling to filter the valueless ones, and applied data extraction to obtain relevant data items.} In the last step, the Constant Comparison method~\cite{glaser1965the} was employed to analyze the extracted data and answer the research questions.

\textbf{Our findings show that}: (1) \textit{Operation Issue} and \textit{Compatibility Issue} are the most common problems faced by developers, (2) \textit{Copilot Internal Error}, \textit{Network Connection Error}, and \textit{Editor/IDE Compatibility Issue} are identified as the most frequent causes, and (3) \textit{Bug Fixed by Copilot}, \textit{Modify Configuration/Setting}, and \textit{Use Suitable Version} are the predominant solutions.

The \textbf{contributions} of this work are that:
\begin{itemize}
    \item We provided a two-level taxonomy for the problems of using Copilot in the software development practice, and a one-level taxonomy for the causes of the problems and the solutions to address the problems.
    \item We drew a mapping from the identified problems to their causes and solutions.
    \item We proposed practical guidelines for Copilot users, the Copilot team, and other researchers.
\end{itemize}

%(1) we identified the problems of using Copilot in the software development practice with a curated dataset~\citep{dataset} and provided a two-level taxonomy for these problems; (2) we identified the causes and solutions of the problems, and came up with a taxonomy for these causes and solutions; (3) we provided the mapping relationship \textcolor{blue}{of identified problems to their causes and solutions}; and (4) we discussed the implications based on the research findings \textcolor{blue}{for Copilot users, Copilot team, and other researchers.}

The rest of this paper is structured as follows: Section \ref{sec:methodology} presents the Research Questions (RQs) and research process. Section \ref{sec:results} provides the results and their interpretation. Section \ref{sec:implication} discusses the implications based on the research results. Section \ref{sec:threats} clarifies the potential threats to the validity of this study. Section \ref{sec:relatedWork} reviews the related work. Finally, Section \ref{sec:conclusions} concludes this work along with the future directions.

\section{Methodology}
\label{sec:methodology}
\begin{figure*}[htbp]
	\centering
	\includegraphics[width=\textwidth]{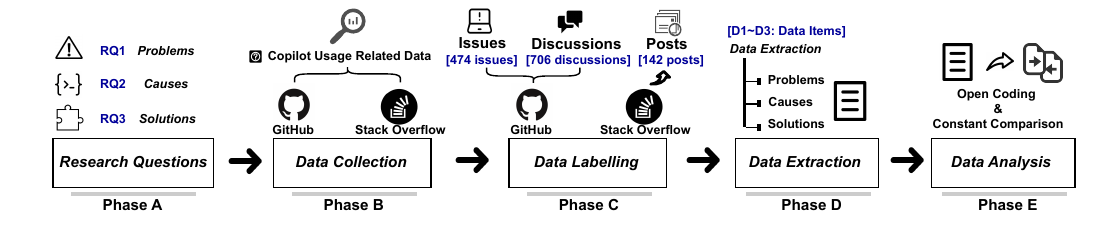}
	\caption{Overview of the research process}
	\label{fig:Overview of research process}
\end{figure*}

The research goal of this study is to identify the problems encountered by users when coding with Copilot, as well as their underlying causes and potential solutions. We formulated three RQs presented in Section \ref{Research Questions}. Fig.~\ref{fig:Overview of research process} provides the overview of the research process.
%\textcolor{blue}{After collecting the data, we conducted data labelling to exclude irrelevant or valueless ones. Subsequently, through data extraction process, we extracted predefined data items from the filtered data, and then analyzed them to obtain the final research results. 

\subsection{Research Questions} 
\label{Research Questions}
\label{RQs}
% Five RQs are formulated according to the goal of this study:\\
\noindent \textbf{RQ1: What are the problems faced by users while using Copilot in software development practice?}\\
\noindent \emph{Rationale}: GitHub Copilot is one of the most popular AI-assisted coding tools and has been widely used for software development with 1.3 million paid users till Feb of 2024~\citep{copilotusers}. Consequently, it is important to understand the specific challenges and problems users face while using this tool in software development practice.\\ %By identifying these problems, this study can help us gain a better understanding of the obstacles that arise when using AI code generation tools like Copilot from the perspective of software developers.\\
\noindent \textbf{RQ2: What are the underlying causes of these problems?}\\
\noindent \emph{Rationale}: Understanding the causes of the problems identified in RQ1 is essential to developing effective solutions to address them. By identifying these causes, this study can provide insights into how to improve the design and functionality of Copilot.\\
\noindent \textbf{RQ3: What are the potential solutions to address these problems?}\\
\noindent \emph{Rationale}:
%RQ3 aims to identify potential solutions to the problems identified in RQ1. 
Exploring the solutions of the problems identified in RQ1 and the causes identified in RQ2 is essential to improving the user experience when using Copilot. By identifying these solutions, this study can gain insight into potential improvements that enhance the functionality and usability of Copilot.\\

%\subsection{Data Collection and Filtering}

%\textcolor{blue}{The research collected data from three sources: GitHub Issues \cite{githubissues}, GitHub Discussions \cite{githubdisucssions}, and Copilot related SO posts. \cite{stackoverflow}} GitHub Issues is a commonly used feature on GitHub for tracking bugs, feature requests, and other issues related to software development projects, which allow us to capture the specific problems and difficulties that users have encountered when coding with Copilot. GitHub Discussions, on the other hand, is a newer feature for more open-ended discussions among project contributors and community members, which also offers a central hub for project-related discussions and knowledge sharing among community members. Discussion topics can cover a wide range from technical questions and proposals to broader topics related to Copilot. \textcolor{blue}{Stack Overflow is a popular technology community that provides a Q\&A platform covering a wide range of programming, development, and technical domains. On this platform, we can access numerous real-world issues encountered by users while using Copilot in their posts.}

\subsection{Data Collection}
We collected data from three sources: GitHub Issues\footnote{\url{https://docs.github.com/en/issues}}, GitHub Discussions\footnote{\url{https://github.com/orgs/community/discussions/categories/copilot}}, and SO posts\footnote{\url{https://stackoverflow.com/}}.  Data from these three sources has been frequently used in empirical studies to gather various developer-reported problems in software engineering \citep{bird2022taking} \citep{musenga2023characterizing} \citep{aoun2021understanding}.
GitHub Issues is a commonly used feature on GitHub for tracking bugs and feature requests and reporting other issues related to software projects. It allows us to capture the specific problems that users have encountered when coding with Copilot.
GitHub Discussions is a feature that allows open-ended discussions among project contributors and community members. It offers a central hub for project-related discussions and knowledge sharing.
% The topics at GitHub Discussions can vary from technical questions and suggestions to usage issues associated with Copilot. 
Stack Overflow is a public Q\&A platform covering a wide range of programming, development, and technology topics, also including inquiries about using Copilot.

Considering that Copilot was announced and started its technical preview on June 29, 2021, we chose to collect the data that were created after that date. The data collection was conducted on June 18, 2023.
To answer RQ3, i.e., the solutions for addressing Copilot usage problems, we chose to collect closed GitHub issues, as well as answered GitHub discussions and SO posts.
%which contain known causes and solutions. As for SO posts, we found that the number of Copilot-related posts was relatively low. Hence, we chose to retrieve all potentially relevant posts in order to obtain a more comprehensive dataset.
Specifically, for GitHub issues, we used ``Copilot'' as the keyword to search closed Copilot-related issues globally in the entire GitHub, and a total of 4,057 issues were retrieved. We also employed ``Copilot'' as a keyword to search answered posts in SO, resulting in 679 retrieved posts. Note that we did not use the ``Copilot'' tag for retrieval because the keyword-based method allows us to obtain a more exhaustive dataset. 
% Different from GitHub issues and SO posts, GitHub discussions are organized into specific subcategories, with ``Copilot'' included as a subcategory under the overarching ``Product'' category. 
Different from GitHub issues and SO posts, GitHub discussions related to Copilot are grouped under the ``Copilot'' subcategory within the ``Product'' category in the GitHub Community, which is recommended by the Copilot team to provide feedback on Copilot usage. Therefore, the discussions within the ``Copilot'' subcategory frequently receive attention and responses from the Copilot team, making them highly valuable for answering RQ2 and RQ3 in our research. Given the high relevance of these discussions to Copilot usage, we collected all the 925 answered discussions under the ``Copilot'' subcategory.

\subsection{Data Labelling}
% To ensure that the data can be used to answer our RQs, we set the filtering criterion: the issue, discussion, or post should contain specific information related to the use of \textcolor{blue}{GitHub} Copilot. We conducted the data labelling on the collected data to filter out the data which cannot be used for this study by following the criterion.
We conducted the data labelling on the collected data to filter out those cannot be used for this study. The criteria for data filtering is as follows: the issue, discussion, or post should contain specific information related to the use of GitHub Copilot.

\subsubsection{Pilot Data Labelling}
To minimize personal bias in the formal labelling process, the first and third authors conducted a pilot data labelling. For GitHub issues and discussions, we randomly selected 100 and 25 from each, making up 2.5\% of the total count. Due to the small quantity of SO posts, we randomly selected 35, which constitutes 5\% of the total posts. Selecting a certain proportion of data from different platforms respectively is to verify whether the criteria of the two authors are consistent across various data sources. The inter-rater reliability between the two authors was measured by the Cohen's Kappa coefficient \citep{jacob1960coefficient}, resulting in values of 0.824, 0.834, and 0.806, which indicate a reasonable level of agreement between the two authors. For any discrepancies in the results, the two authors engaged in discussions with the second author to reach a consensus. To be specific, the first author and the third author presented their reasons for why the data should or should not be included in our research. Then, the three authors analyzed and discussed these reasons according to the criteria for data filtering, ultimately arriving at a conclusion accepted by the three authors. The results of pilot data labelling were compiled and recorded in MS Excel \citep{dataset}.

\subsubsection{Formal Data Labelling}
The first and third authors then conducted the formal data labelling. During this process, we excluded a large amount of data not related to our research. For instance, ``Copilot'' may refer to other meanings in some situations, such as the ``co-pilot'' of an aircraft. Additionally, Copilot might be mentioned in a straightforward manner without additional information, like a post mentioned, ``\textit{You can try using Copilot, which is amazing}''. We also excluded such cases of data since they could not provide useful information about the usage of Copilot. During the labelling process, any result on which the two authors disagreed was subject to discussion with the second author until an agreement was reached. Ultimately, the two authors collected 473 GitHub issues, 706 GitHub discussions, and 142 SO posts. The data labelling results were compiled and recorded in MS Excel \citep{dataset}. 

\subsection{Data Extraction} 
To answer the three RQs in Section \ref{Research Questions}, we established a set of data items for data extraction, as presented in Table~\ref{Data Items and RQ}. Data items D1-D3 intend to extract the information of problems, underlying causes, and potential solutions from the filtered data to answer RQ1-RQ3, respectively. These three data items could be extracted from any part of a GitHub issue, discussion, or SO post, such as the title, the problem description, comments, and discussions.

\subsubsection{Pilot Data Extraction}
The first and third author conducted a pilot data extraction on 20 randomly selected GitHub issues, 20 discussions, and 20 SO posts, and in case of any discrepancies, the second author was involved to reach a consensus. The results indicated that the three data items could be extracted from our dataset. Based on the observation, we established the following criteria for formal data extraction: (1) If the same problem was identified by multiple users, we recorded it only once. (2) If multiple problems were identified within the same GitHub issue, GitHub discussion, or SO post, we recorded each one separately. (3) For a problem that has multiple causes mentioned, we only recorded the cause confirmed by the reporter of the problem or the Copilot team as the root cause. (4) For a problem that has multiple solutions suggested, we only recorded the solutions that were confirmed by the reporter of the problem or the Copilot team to actually solve the problem. 

\subsubsection{Formal Data Extraction}
\label{sec:formalDataExtraction}
The first and third authors conducted the formal data extraction from the filtered dataset to extract the data items. Subsequently, they discussed and reached a consensus with the second author on inconsistencies to ensure that the data extraction process adhered to the predetermined criteria. Each extracted data item was reviewed multiple times by the three authors to ensure accuracy. The data extraction results were compiled and recorded in MS Excel \citep{dataset}. 

\begin{table}[htbp]
\scriptsize
\caption{Data items extracted and their corresponding RQs}
\label{Data Items and RQ}
\begin{tabular}{m{0.1cm}<{\centering}m{1.21cm}m{5.2cm}m{0.3cm}<{\centering}}
\hline
\textbf{\#} & \textbf{Data Item}    & \textbf{Description}                                                     & \textbf{RQ}   \\ \hline
D1          & Problem    & \textit{The key point(s) of the problem from GitHub issues, GitHub discussions, and SO posts}           & RQ1   \\ \hline
D2          & Cause    & \textit{The key point(s) of the cause from GitHub issues, GitHub discussions, and SO posts}             & RQ2   \\ \hline
D3          & Solution & \textit{The key point(s) of the solution from GitHub issues, GitHub discussions, and SO posts}          & RQ3   \\ \hline
\end{tabular}
\end{table}

% Note that not all collected data includes the cause and solution of a problem. Although we selected closed GitHub issues, answered GitHub discussions and SO posts, the specifics of each piece of data vary significantly.
Although we selected closed GitHub issues, and answered GitHub discussions and SO posts, not all collected data includes the cause and solution of a problem. Sometimes, the respondents to a Copilot related problem might offer a solution without detailed analysis, preventing us from extracting the underlying causes. In other situations, although the cause of a problem was identified, the user did not describe the specific resolution process. For example, a user found that Copilot ``\textit{cannot work correctly on VSCode remote server}'' and realized it was due to ``\textit{the bad network}'', but did not provide any solutions (Discussion \#14907). Additionally, even when some responses provided both causes and solutions, they might not be accepted or proven effective by the problem's reporter or the Copilot team. For example, a user asked for ``\textit{a way to set up GitHub copilot in Google Colab}'', but the user neither accepted nor replied to the three proposed answers (SO \#72431032). Therefore, we cannot consider any of the three answers as an effective solution to his problem.

\subsection{Data Analysis}
%In order to address the three RQs in Section \ref{sec:mapping}, we conducted data analysis using the Constant Comparison method \cite{glaser1965the}. This analytical approach assists us to identify patterns and differences by comparing the data, leading to the formation of higher-level categories. Firstly, the first author carefully reviewed each extracted data item and conducted initial coding for issues, causes, and solutions. Then, the first author compared and merged all the coded results to form a series of types and categories. We repeated this process to ensure the accuracy and reliability of the final results. Ultimately, the first author discussed and refined the results with the second and third authors. 

\begin{figure*}[htbp]
	\centering
	\includegraphics[width=0.8\linewidth]{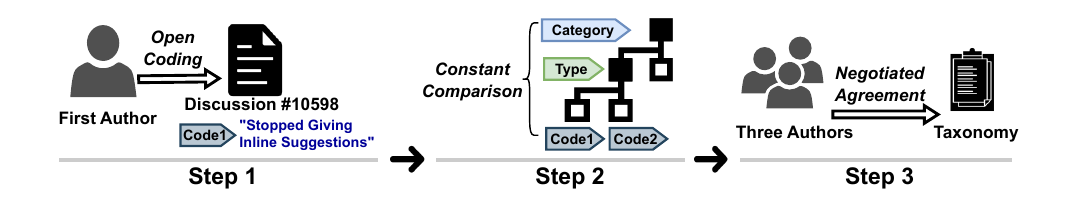}
	\caption{The process of data analysis}
	\label{fig:The process of data analysis}
\end{figure*}

To answer the three RQs formulated in Section \ref{sec:methodology}, we conducted data analysis by using the Open Coding and Constant Comparison methods, which are two widely employed techniques from Grounded Theory during qualitative data analysis \citep{stol2016grounded}. Open Coding is a research method that encourages researchers to generate codes based on the actual content of the data rather than being restricted by pre-existing theoretical frameworks. These codes constitute descriptive summarizations of the data, aiming to capture the underlying themes. For Constant Comparison, researchers continuously compare the coded data, dynamically refining and adjusting the categories based on their similarities and differences before forming the final classification results.

As presented in Fig \ref{fig:The process of data analysis}, the specific process of data analysis includes three steps: 1) The first author reviewed the collected data and then assigned descriptive codes that succinctly encapsulated the core themes. For instance, the issue in Discussion \#10598 was coded as ``\textit{Stopped Giving Inline Suggestions}'', which was reported by a user whose Copilot suddenly stopped providing code suggestions in VSCode. 2) The first author compared different codes to identify patterns, commonalities, and distinctions among them. Through this iterative comparison process, similar codes were merged into higher-level types and categories. For example, the code of Discussion \#10598, along with other akin codes, formed into the type of \textsc{functionality failure}, which further belongs to the category of \textit{Operation Issue}. Once uncertainties arose, the first author engaged in discussions with the second and third authors to achieve a consensus. Due to the nature of Constant Comparison, both the types and the categories were refined multiple times before reaching their final form. 3) The initial version of the analysis results was further verified by the second and third authors, and the negotiated agreement approach~\citep{campbell2013coding} was employed to address the conflicts. The final results are presented in Section \ref{sec:results}.

\section{Results and Interpretation}
\label{sec:results}
In this section, we report the results of three RQs and provide their interpretation. 
%In Section \ref{subsec:issuesresults}, we presented the types of problems, while in Sections \ref{subsec:causesresults} and \ref{subsec:solutionsresults}, we present the types of causes and solutions for the corresponding issues, respectively. 
The results of Copilot usage problems are categorized into two levels: categories (e.g., \textit{Suggestion Content Issue}) and types (e.g., \textsc{less efficient suggestion}). Meanwhile, the results for causes and solutions are organized as types only (e.g., \textit{Network Connection Error}). We also provide the mapping relationship of Copilot related problems to their causes and solutions. As mentioned in Section~\ref{sec:formalDataExtraction}, only causes that were proven to lead to the problems and solutions that could resolve the problems were extracted and provided in the results. Therefore, not all problems have corresponding causes and solutions. It is worth noting that due to the rapid update of Copilot, some problems and feature requests raised by users have already been addressed in newer releases of Copilot. We identified these two scenarios separately as two types of solutions, i.e., \textit{Bug Fixed by Copilot} and \textit{Feature Implemented by Copilot}. However, due to the absence of Copilot version information in the dataset, we could not consider the version information thoroughly in this study. To help understand the taxonomies of problems, causes, and solutions of using Copilot, we provide examples with the ``\#'' symbol, which indicates the ``GitHub Issue ID'', ``GitHub Discussion ID'', or ``SO Post ID'' in the dataset~\citep{dataset}.

\begin{figure*}[htbp]
	\centering
	\includegraphics[width=\linewidth]{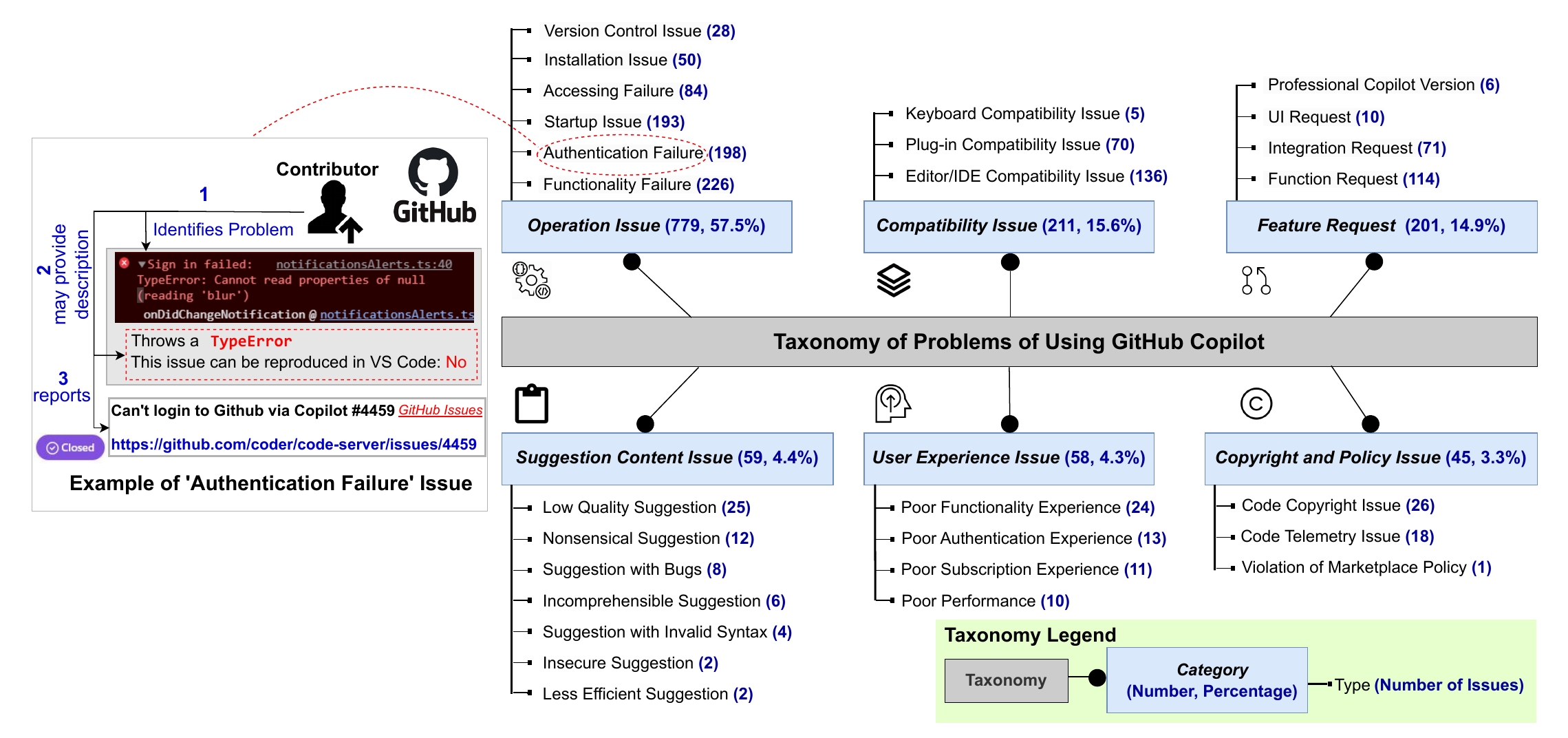}
	\caption{A taxonomy of problems when using GitHub Copilot}
	\label{fig:Results of RQ1}
\end{figure*}

\subsection{Type of Problems (RQ1)}
\label{subsec:issuesresults}
%\noindent \textbf{RQ1: What are the issues faced by users while using Copilot in practice software development?}\\
Fig.~\ref{fig:Results of RQ1} presents the taxonomy of the problems extracted from our dataset. We have identified a total of 1,355 problems related to Copilot usage. It can be observed that \textit{Operation Issue} (57.5\%) accounts for the majority of problems faced by Copilot users. Furthermore, there are a notable number of users who have countered \textit{Compatibility Issue} (15.6\%) when using Copilot in different environments, followed by users who have raised \textit{Feature Request} (14.9\%) based on their user experience and requirements. Additionally, smaller percentages were identified as \textit{Suggestion Content Issue} (4.4\%), \textit{User Experience Issue} (4.3\%), and \textit{Copyright and Policy Issue} (3.3\%).
% Besides, a substantial number of users have raised \textit{Feature Request} (15.0\%) based on their user experience and requirements. There is also a portion of users who have encountered \textit{Compatibility Issue} (15.6\%) when using Copilot in different environments, while smaller percentages were identified as \textit{Suggestion Content Issue} (4.4\%), \textit{User Experience Issue} (4.3\%), and \textit{Copyright and Policy Issue} (3.3\%).

\subsubsection{\textit{Operation Issue (57.5\%)}} Operation Issue refers to a category of obstacles encountered by users when attempting to utilize some of the fundamental functions of Copilot. This category of problems is divided into six types, which are elaborated below.
\begin{itemize}
    \item  \textsc{functionality failure} refers to the abnormality of code generation-related features provided by Copilot. Copilot offers various interactive features to better engage with users, such as ``previous/next suggestion'', ``viewing all suggestions'', and ``configuration of shortcut keys to accept suggestions''. Users may encounter exceptions when using these features. For example, a user reported that ``\textit{Copilot no longer suggesting on PyCharm}'' after a period of not using it (Discussion \#11199).
    % \item  \textsc{setup/operation issue} refers to errors or malfunctions that occur during the initialization or operation of Copilot, and often involve runtime exceptions. These issues can prevent Copilot from running correctly, or cause it to crash unexpectedly, such as when a user encounters ``\textit{copilot start error in VSCode}'' (Discussion \#30996).
    \item \textsc{startup issue} refers to errors or malfunctions encountered by users attempting to run Copilot. This issue results in a complete failure of Copilot to execute and is typically accompanied by error messages. Such problems may arise either during a user's initial usage of Copilot or unexpectedly after several successful runs. For example, a user failed to activate Copilot after installing it on VSCode, and received an error message stating ``\textit{Cannot find module}'' (Issue \#380).
    \item  \textsc{authentication failure} refers to the issues related to user login and authentication difficulties when using Copilot. Copilot requires users to log in to their GitHub account before using the service. Only users with access permissions (including paid subscriptions, student identity verification, etc.) can use the code generation service of Copilot. During the authentication process, users may encounter \textsc{authentication failure}, resulting in the inability to use Copilot. 
    % For example, a user mentioned in the discussion forum that ``\textit{I cannot log in after the upgrade}'' (Discussion \#18132). 
    For example, a user was repeatedly prompted with the message ``\textit{waiting for GitHub Authentication}'' in IDEA, and was unable to log in (SO \#72505280).
    \item  \textsc{accessing failure} refers to the situation where users fail to access Copilot's server, which often involves errors related to server connections. A user may encounter an error message like ``\textit{GitHub Copilot could not connect to server}'' (Discussion \#11801). 
    \item  \textsc{installation issue} refers to the problems encountered by users during the installation process of Copilot, including installation errors, inability to find installation methods, and other related problems. 
    % For instance, some users may encounter issues such as ``\textit{Errors when installing Copilot}'' (Discussions \#17250).
    For instance, a user failed to install Copilot on VSCode insiders, and the server log showed ``\textit{Error while installing the extension}'' (Issue \#187).
    \item  \textsc{version control issue} refers to the problems that users encounter when adjusting the version of Copilot or its runtime environment (e.g., IDE), including the inability to upgrade the Copilot version or abnormal issues like continuing to prompt for upgrades even after upgrading. For example, a user reported that ``\textit{copilot plugin fails to update}'' when using it in IntelliJ IDEA (Discussion \#17298).
\end{itemize}

\textbf{Interpretation}: We identified \textit{Operation Issue} at various stages of user interaction with Copilot. Users tend to report these problems and seek assistance, making \textit{Operation Issue} the most prevalent category of problems related to Copilot. \textsc{functionality failure} (226), \textsc{authentication failure} (198),  and \textsc{startup issue} (193), are the top three types of such problems. We attribute the higher frequency of the first two types to the deficiencies in Copilot's feature design and stability, which are also influenced by users' environments in which Copilot operates. \textsc{authentication failure} mainly stems from particular details encountered during the login process when users need to access Copilot with their GitHub accounts.

\subsubsection{\textit{Compatibility Issue (15.6\%)}}This category covers the problems that arise from mismatches between Copilot and its runtime environment.  Copilot operates as a plugin in various IDEs and code editors (e.g., VSCode and IntelliJ IDEA), and the complexity of the environments Copilot operates on can result in an increased number of compatibility issues. These problems are further classified into three types, which are elaborated below.
\begin{itemize}
    \item  \textsc{editor/ide compatibility issue} refers to issues arising from mismatches between Copilot and its IDE or editor. These problems typically manifest as Copilot being unable to operate properly in a specific IDE or editor. For example, a user previously found that Copilot ``\textit{does not work in Neovim}'' while writing a Python program, even though the Copilot status showed that ``\textit{Copilot: Enabled and Online}'' (SO \#72174839).
    \item  \textsc{plug-in compatibility issue} refers to a type of matching issue that arises when Copilot and other plugins are activated and working together at the same time. Such problems can cause partial or complete malfunctions of Copilot and other plugins. They are usually identified through troubleshooting methods such as disabling Copilot or other plugins. For instance, one user reported that ``\textit{a Keyboard shortcut conflict with Emmet}''  prevented him from receiving code suggestions generated by Copilot (Issue \#47).
    % \item  \textsc{framework compatibility issue} refers to a type of compatibility problem between Copilot and the framework it operates on. One common example is the compatibility issue between Copilot.vim \citep{copilotvim}, an official version of Copilot designed specifically for Vim, and Node.js. For example, a user updated his Copilot.vim but got a message that ``\textit{the current version of Node.js 18.2.0 is no longer supported}'', although Copilot even worked fine in an earlier version (Discussion \#16800).
    \item  \textsc{keyboard compatibility issue} refers to the situation when the functionality of Copilot can not be used in some uncommon keyboard layouts. For example, a user with a German keyboard layout could not use most of the code generation-related features of Copilot (Discussion \#7094).
\end{itemize}

\textbf{Interpretation}: \textit{Compatibility Issue} arises from the complex environments in which users utilize Copilot, as well as the compatibility robustness of Copilot itself. In the case of \textsc{editor/ide compatibility issue} (136), VSCode, the platform officially recommended for using Copilot, has garnered a higher number of reported issues about compatibility. We also identified many problems in other widely used IDEs, like Visual Studio, IntelliJ IDEA, and PyCharm. The appearance of \textsc{plug-in compatibility issue} (70) is less predictable, which often arises when using Copilot with other code completion tools.
 
\subsubsection{\textit{Feature Request (14.9\%)}} Feature Request refers to the features that users request to add or improve based on their experience and actual needs when using Copilot. These feature requests not only help improve the user experience of Copilot but also contribute to the exploration of how AI code generation tools like Copilot can better interact with developers. This category is further divided into four types, as shown below. 
\begin{itemize}
    \item  \textsc{function request} refers to the requests for developing new functions in Copilot, which typically arise from users' genuine needs and difficulties encountered while utilizing the tool. 
    % For example, a user has suggested that the addition of a ``\textit{Code Explanations Feature}'' could enhance the usefulness of Copilot (Discussion \#7509).
    For example, a user requested that Copilot should be able to ``\textit{look at the context of a project with multiple files}'', rather than generate code suggestions based on the context in a single file (SO \#73848372).
    \item  \textsc{integration request} refers to a type of request for Copilot to be available on certain platforms or to be integrated with other plugins. This is mainly due to the desire of some users to use Copilot in the environments they are familiar with. For instance, a user called for ``\textit{Support for Intellij 2022.2 EAP family}'' (Discussion \#17045). The requests for integration also reflect the popularity of Copilot among developers to some extent.
    \item  \textsc{ui request} refers to requests made by users for changes to the User Interface (UI) of Copilot, such as modifying the appearance of the Copilot icon. These requests generally aim to improve the visual effects and user experience of Copilot. For example, a user requested the addition of a ``\textit{status indicator}'' to provide information about the current working status of Copilot (Issue \#161).
    \item  \textsc{professional copilot version} refers to requests from some users for a professional version of Copilot. These users are developers from certain companies who hope to receive more professional and reliable code generation services in their actual work. They may have higher requirements for the reliability and security of Copilot's code, as well as team certification and other aspects. For example, a user asked for ``\textit{an on-prem version for companies to purchase}'', so that they could deploy Copilot in a local environment (Discussion \#38858).
\end{itemize}

\textbf{Interpretation}: For \textsc{function request} (114), we observed that users expressed a desire for greater flexibility in configuring Copilot to align more closely with their development habits. Common requests include the ability to accept Copilot's suggestions word by word and to specify where Copilot should automatically operate in terms of file types or code development scopes. More innovative demands involve the need for Copilot to provide suggestions according to the whole project, as well as features like code explanation and chat functionality which have been provided in GitHub Copilot Chat~\citep{copilotchat}. \textsc{integration request} (71) reflects the wish of developers to use Copilot in their familiar environments. This places greater demands on the Copilot team, as we have identified a significant number of \textit{Compatibility Issues}.

\subsubsection{\textit{Suggestion Content Issue (4.4\%)}} This category of problems refers to the issues related to the content of the code generated by Copilot. The generation of code suggestions is the core feature of AI code generation tools like Copilot, and the quality of the suggestions directly determines whether users will adopt them. Therefore, the content of the generated code is naturally an area of concern for users, researchers, and the Copilot team. These problems are further divided into seven specific situations, which are detailed below.
\begin{itemize}
    \item \textsc{low quality suggestion} refers to situations where Copilot is unable to comprehend the context sufficiently to generate useful code. Such code suggestions may not have any syntactical errors, but due to their poor quality, they are unlikely to be adopted by users. For instance, Copilot once generated an empty method containing only a return statement without meeting the requirements specified in the user's code (Discussion \#6631).
    \item \textsc{nonsensical suggestion} refers to the code suggestions provided by Copilot that are completely irrelevant to the needs of users. Such suggestions are considered almost unusable and provide little heuristic assistance to the user. For example, a user once received an inaccessible fake URL generated by Copilot, which was of no help with his programming task (Discussion \#14212). 
    \item \textsc{suggestion with bugs} refers to the situation where Copilot is able to generate relevant code based on the context, but the suggested code contains some bugs. This can result in the program being able to run, but not as the developer intended, or in some cases, it may cause errors or crashes. For example, a user reported that Copilot suggested using ``\textit{setState(!state)}'' instead of ``\textit{setState(true)}'', which caused a logical bug in his code (Issue \#43).
    \item \textsc{incomprehensible suggestion} refers to the situation where Copilot provides potentially useful code suggestions, but due to the complexity of the code or user's lack of experience, they found it challenging to comprehend the suggested code and need more time to verify its correctness. For example, a user complained that ``\textit{My Github Copilot just autocompleted it for me, then I scoured the internet trying to find information pertaining to it but could not}'' (SO \#73075410).
    \item \textsc{suggestion with invalid syntax} refers to the situation where the suggestions generated by Copilot may contain syntax errors that prevent the program from running properly. For example, a user found that the code suggestions provided by Copilot ``\textit{missed curly brackets, leading to erroneous code when accepting suggestion}'' (Discussion \#38941).
    % One example is when the suggested code is missing a closing bracket, causing the editor to display a syntax error (SO \#74511049).
    % One example is when a user was writing C code, and the suggestion provided by Copilot was missing a closing bracket, resulting in a syntax error being displayed in his editor (SO \#74511049).
    \item  \textsc{less efficient suggestion} refers to the code suggestions generated by Copilot that are functionally correct and meet the requirements of users, but may suffer from suboptimal execution efficiency or convoluted logic, potentially impacting the overall quality of the code. For example, when a user requested Copilot to ``\textit{find the cup with the most water}'' and ``\textit{find the maximum amount of water in any cup}'', the suggested code performed adequately but was not optimized for efficiency (Issue \#160).
    \item  \textsc{insecure suggestion} refers to the code suggestions generated by Copilot that introduce security vulnerabilities. For example, a user indicated that the code suggestion lacked accountability for the sizes being read (Discussion \#6636).
\end{itemize}

\textbf{Interpretation}: The quality of code suggestions is a critical factor in determining the capability of Copilot for practical code development. We identified a relatively small number of \textit{Suggestion Content Issues}, possibly indicating that users are less inclined to report problems related to suggested code compared to usage-related problems. Among these problems, \textsc{low quality suggestion} (25), \textsc{nonsensical suggestion} (12), and \textsc{suggestion with bugs} (8) are the three most frequently reported types, while \textsc{insecure suggestion} (2) and \textsc{less efficient suggestion} (2) are less prevalent. This result implies that the main concern for users could be whether Copilot can provide code suggestions that have significant referential value.

\subsubsection{\textit{User Experience Issue (4.3\%)}} This category covers user feedback on their experience of using Copilot. Compared with \textit{Operation Issue}, Copilot generally runs and functions as intended, but the user experience is suboptimal. \textit{User Experience Issue} can provide insights into areas where Copilot could be improved. \textit{User Experience Issue} can be further classified into four types, which are detailed below.
\begin{itemize}
    \item \textsc{poor functionality experience} refers to a type of user experience problem where the usage of Copilot's code generation-related functionalities is unsatisfactory. When such problems arise, Copilot's functionalities remain operational, which is different from \textsc{functionality failure}. However, users expressed dissatisfaction with their experience when using these functionalities. These problems can often hinder the coordination between users and Copilot, and even decrease the efficiency of development work. Such problems often highlight areas where Copilot could be further enhanced and may potentially motivate users to propose some \textit{Feature Requests}.
    % For instance, a user complained that the automatically generated suggestions provided by Copilot were highly distracting, forcing him to manually trigger the code generation functionality (Discussion \#13007).
    For example, a user felt that the code automatically generated and popped up by Copilot was quite noisy (Issue \#97).
    \item \textsc{poor subscription experience} refers to the obstacles that users encounter during the process of subscribing to the services of Copilot. Copilot offers several subscription methods (e.g., student verification, paid subscription), leading to some inconvenience for users during this process. For example, one user felt lost and was unsure about what to do next after setting up a billing (Discussion \#19119).
    \item \textsc{poor performance} refers to performance issues that occur when Copilot is running, which directly impacts the user experience. These problems include high CPU usage, long response times, and overly frequent server access. For example, a user complained that Copilot took ``\textit{around 1-2 minutes for it to show one suggestion}'' on VSCode, which was very slow (Discussion \#19491).
    \item \textsc{poor authentication experience} refers to the inconvenience that users encounter when authenticating their identities before using Copilot. While users successfully navigate the login procedure, they may experience a suboptimal user experience. This could be due to factors such as an unwieldy process flow or the absence of explicit instructions. For example, a user complained that Copilot frequently ``\textit{prompt to enable GitHub Copilot on every VSCode launch}'' which can be a significant source of frustration (SO \#70065121).
\end{itemize}

\textbf{Interpretation}: \textit{User Experience Issues} provide valuable insights into the direction for improving Copilot. Among the \textsc{poor functionality experience} (24), the most commonly reported problems involve Copilot's inline suggestions that cause disruptions to the coding process of users (5) and the inconvenience of not being able to accept certain portions of the suggested code (2). These concerns align with some of the demands mentioned by users in \textit{Feature Request}, e.g., setting when Copilot can generate code and the length of suggested code.

\subsubsection{\textit{Copyright and Policy Issue (3.3\%)}} Copilot is trained on a large corpus of open source code and generates code suggestions based on the users' code context. The way in which Copilot operates raises concerns regarding potential copyright and policy issues, as expressed by some users. These problems are divided into three types, as shown below.

\begin{itemize}
    \item \textsc{code copyright issue} refers to the concerns raised by some code authors regarding the unauthorized use of their open-source code by Copilot for model training. GitHub is currently one of the most popular web-based code hosting platforms, and since the release of Copilot, there have been suspicions among some code authors that their code hosted on GitHub has been used for training without proper consideration of their license. For example, a user ``\textit{started migration}'' of his projects on GitHub because he was worried about his code being used to train Copilot without permission (Issue \#148).
    \item \textsc{code telemetry issue} refers to the concerns expressed by users regarding Copilot collecting their code to generate suggestions, which may potentially result in the leakage of confidential code. Some users may also simply be unwilling to have their own code, as well as the code generated by Copilot for them, collected for other purposes. For example, a user is concerned that using Copilot might ``\textit{give away API key}'' in his code (SO \#70559637).
    \item  \textsc{violation of marketplace policy} is a specific case where a user reported that Copilot was able to be published on the VSCode marketplace despite using proposed APIs, while other plugins were prohibited. The user suspected that this behavior may be in violation of the Marketplace Policy (Issue \#3).
\end{itemize}

\textbf{Interpretation}: The emergence of \textit{Copyright and Policy Issue} reveals the concerns of users about the way Copilot works. Copilot is trained on multi-language open-source code and also needs to collect users' code context during its operation to generate suggestions. These two facts have led people to pay more attention to copyright and intellectual property problems when using Copilot, especially in in-house development.

\subsection{Type of Causes (RQ2)}
\label{subsec:causesresults}
%\noindent \textbf{RQ2: What are the underlying causes of these issues?}\\
\subsubsection{\textbf{Results}} As mentioned in Section \ref{sec:formalDataExtraction}, not all problems have corresponding causes that can be extracted. As a result, we identified a total of 391 causes, which were collected from 28.9\% of all problems related to Copilot usage, and categorized into 16 types as presented in Table \ref{The Types of causes leading to the issues when using Copilot}. The result indicates that the most frequent causes are \textit{Copilot Internal Error} (19.4\%) and \textit{Network Connection Error} (13.6\%), with \textit{Editor/IDE Compatibility Issue} (12.8\%) and \textit{Unsupported Platform} (8.2\%) also commonly reported. The example, count, and proportion of each type of cause are presented in Table \ref{The Types of causes leading to the issues when using Copilot}. 
%Due to the space limit, we interpret the top five most frequent causes. 
It is worth noting that certain types of problems can potentially be the causes of other problems. For example, \textsc{editor/ide compatibility issue} is a type of \textit{Compatibility Issue}. However, it is also the causes leading to other types of problems such as \textsc{installation issue} and \textsc{startup issue}.
% If a compatibility issue with an IDE or editor is brought up by the presenter when describing the problem initially, we categorize it as a type of problem. If the compatibility issue is only raised by others later on in the discussion, we classify it as one of the causes contributing to other functional abnormalities.}

\begin{table*}[htbp]
\renewcommand{\arraystretch}{1.4}
\scriptsize
% \small
\caption{Causes of Copilot usage problems}
\label{The Types of causes leading to the issues when using Copilot}
    
%\begin{tabular}{m{3.35cm}m{11.9cm}m{0.6cm}<{\centering}m{0.6cm}<{\centering}}
\begin{tabular}{m{5.2cm}m{9cm}m{0.8cm}<{\centering}m{0.8cm}<{\centering}}
\hline
\textbf{Type}                   & \textbf{Example (Extracted Cause)}                                                                      & \textbf{Count}   & \textbf{\%}           \\ \hline
Copilot Internal Error (CIE)         & \textit{There has been an outage of one of our models which has caused the others to have to handle a higher traffic load.} (Discussion \#14370)                                                   & 76               & 19.4\% \\ \hline
 Network Connection Error (NCE)       & \textit{It seems that my company's network settings blocked the connection to copilot and that affect the behavior of my personal mac too.} (Discussion \#36152)        & 53               & 13.6\% \\ \hline
 % Editor/IDE Compatibility Issue  & \textit{this was caused by a bug, which did not properly terminate the Copilot helper process when the IDE terminated.} (Discussion \#17298)                          & 37               & 11.1\% \\ \hline
 Editor/IDE Compatibility Issue (EICI) &  \textit{This is related to the version of your visual studio 2022.} (SO \#71702171)                          & 50               & 12.8\% \\ \hline
% Unsupported Platform            & \textit{Sadly the extension does not work with VSCodium at the moment.} (Discussion \#19726)     & 31               & 9.2\% \\ \hline
Unsupported Platform (UP)           & \textit{$\cdots$ as some Linux distributions ship with Code OSS or VSCodium which Copilot does not currently support.} (Discussion \#8015)     & 32               & 8.2\% \\ \hline
% Improper Configuration/Setting & \textit{my ``Inline Suggestions'' were turned off in settings.} (Discussion \#10598)   & 27               & 8.0\% \\ \hline
Improper Configuration/Setting (ICS) & \textit{The settings.json file had disabled inline suggestions.} (SO \#76257401)   & 27               & 6.9\% \\ \hline
Poor Functionality Experience (PFE) & \textit{Please add shortcut command for suggestion ... Feature auto suggestion is something annoying.} (Discussion \#7172)   & 26               & 6.6\% \\ \hline
For Coding Habit (FCH) & \textit{I would like to change the Tab keybinding for completion in vim to something else because many people use the tab binding for opening the completion menu.} (Discussion \#6919)   & 25               & 6.4\% \\ \hline
 User Unauthorized (UU)              & \textit{I'm sorry that you have not been given access yet.} (Discussion \#16795) & 22               & 5.6\% \\ \hline
Improper User Operation (IUO)             & \textit{I had signed up to Copilot under a different GitHub account} (Discussion \#19556)      & 17               & 4.3\% \\ \hline
% Framework Compatibility Issue (FCI) & \textit{We believe this was probably because of a broken Node installation which we download to run Copilot on. } (Issue \#337)             & 16                & 4.1\% \\ \hline
Plug-in Compatibility Issue (PCI)  &\textit{This was a bad extension interaction between Copilot and learn-markdown.} (Issue \#438) & 15  & 3.8\% \\ \hline
%Obsoleted IDE/Editor Version    &\textit{you need to update VSCode then newer extensions will be available.} (Discussion \#34627)   & 16    & 4.7\% \\ \hline
Intentional Design of Copilot (IDC)  &\textit{Yes, this is intentional. There is a hidden setting that you can toggle to disable this behavior.} (Issue \#152)  & 12    & 3.1\% \\ \hline
% Unimplemented Feature           &\textit{Currently the Copilot VSCode extension does not support proxies.} (Discussion \# 11630)  & 14             & 4.2\% \\ \hline
Unimplemented Feature (UF)         & \textit{... currently the Copilot VSCode extension does not support proxies.} (Discussion \#11630)  & 12             & 3.1\% \\ \hline
% Plug-in Compatibility Issue   &\textit{This is usually interference with markdown highlighting extensions.} (Discussion \#13668) & 14  & 4.2\% \\ \hline
For Higher Coding Efficiency (FHCE)  &\textit{Suggestion: Add mapping generator for classes ... Would be nice to have such feature, that would help with every day mapping.} (Discussion \#7870) & 9  & 2.3\% \\ \hline
Obsoleted Copilot Version (OCV)      &\textit{Your copilot extension is out of date} (Discussion \#17463)  & 7  & 1.8\% \\ \hline
License Restriction (LR)            & \textit{Due to the licensing of Copilot, it cannot be used in free or open-source software such as code-server} (Issue \#122) & 4     &1.0\% \\ \hline
Code Telemetry Issue (CTI)            & \textit{I'd like to disable copilot per workspace, so I can use it for open-source projects but not private/work projects.} (Discussion \#47991) & 4     &1.0\% \\ \hline
%Unclear Instruction             & \textit{The code is displayed in your IDE or editor, it's not a two-factor code.} (Discussion \#8468)    & 3  & 0.9\% \\ \hline
    \end{tabular}
\end{table*}

\begin{itemize}
    \item \textit{Copilot Internal Error} (CIE) refers to the problems with the Copilot server side that affects its usage. These may include errors with the CodeX model and services provided by the server side, which are not visible to users. For example, the Copilot team once reported that ``\textit{there has been an outage of one of our models which has caused the others to have to handle a higher traffic load}'', which consequently results in temporary issues (Discussion \#14370).
    \item \textit{Network Connection Error} (NCE) refers to the disruptions in the network communication between the user side and the Copilot server side, resulting in an inability for users to utilize the code generation service of Copilot. For example, a user repeatedly encountered issues with Copilot not functioning properly because his ``\textit{company's network settings blocked the connection to copilot}'' (Discussion \#36152).
    \item \textit{Editor/IDE Compatibility Issue} (EICI) refers to situations where the code editors and IDEs is not compatible with Copilot, resulting in various anomalies when using Copilot. For example, a user was ``\textit{unable to install GitHub Copilot extension in Visual Studio 2022 Enterprise}'', because the version of his IDE was outdated, resulting in its incompatibility with Copilot (SO \#71702171).
    \item \textit{Unsupported Platform} (UP) refers to situations where users try to use Copilot on development platforms that lack official support for Copilot, which may results in unpredictable problems. For example, the Copilot team claimed that ``\textit{Code OSS and VSCodium}'' were not supported by Copilot, which explains why some users failed to install Copilot in these two IDEs (Discussion \#8015).
    \item \textit{Improper Configuration/Setting} (ICS) refers to situations where Copilot operates abnormally or offers a suboptimal user experience due to the settings that are not configured appropriately. For example, a user found that ``\textit{the setting.json file had disabled inline suggestions}'', which resulted in Copilot not providing code suggestions in VSCode (SO \#76257401).
    \item \textit{Poor Functionality Experience} (PFE) refers to the negative experiences users encounter when coding with Copilot, which lead them to request for enhancement to existing features. For example, a user wanted Copilot to provide code suggestions via a shortcut, because ``\textit{feature auto suggestion is something annoying}'' (Discussion \#7172).
    \item \textit{For Coding Habit} (FCH) refers to the wishes of some users for Copilot to offer new features and ensure compatibility with their preferred development platforms to accommodate their coding habits. For example, a user asked ``\textit{if there was an option for changing the keybinding for accepting the suggestions}'', because he was accustomed to ``\textit{using the tab binding for opening the completion menu}'' (Discussion \#6919).
    \item \textit{User Unauthorized} (UU) refers to situations where some users cannot access Copilot services because they lack the authorization for using Copilot. For example, a user found that ``\textit{GitHub Copilot could not connect to server}'', because he was not authorized yet (Discussion \#16795).
    \item \textit{Improper User Operation} (IUO) refers to situations where Copilot exhibits unintended behavior because of user mistakes or oversights during the process of using Copilot such as registration, login, and subscription. For example, a user consistently received a prompt in Visual Studio stating that ``\textit{Your Copilot experience is not fully configured, please complete your setup}''. It was found that this issue arose because he ``\textit{had signed up to Copilot with a different GitHub account}'' (Discussion \#19556).
    % \item \textit{Framework Compatibility Issue} (FCI) refers to situations where Copilot is not compatible with its framework, resulting in it being unable to operate. For example, a user found that Copilot.vim could not be activated ``\textit{because of a broken Node installation}'' for Copilot to operate on (Issue \#337).
    \item \textit{Plug-in Compatibility Issue} (PCI) refers to situations where Copilot fails to work properly because of incompatibility with other plug-ins. For example, ``\textit{a bad extension interaction between Copilot and learn-markdown}''  resulted in some users being unable to accept code suggestions from Copilot (Issue \#438).
    \item \textit{Intentional Design of Copilot} (IDC) refers to situations where what some users may perceive as anomalies of Copilot are features deliberately designed by the Copilot team. For example, a user noticed that Copilot's ``\textit{inline completions are trumping normal completions}''. However, a member from the Copilot team proved that ``\textit{this is intentional}'' and can be disabled by modifying certain setting (Issue \#152).
    \item \textit{Unimplemented Feature} (UF) refers to the functionalities that users assume Copilot to already have, but which the Copilot team has not yet provided. For example, a user could not activate Copilot behind proxy in VSCode, and a Copilot team member explained that ``\textit{currently the Copilot VSCode extension does not support proxies}'' (Discussion \#11630).
    \item \textit{For Higher Coding Efficiency} (FHCE) is the reason that some users wish for new functionalities in Copilot to enhance efficiency in coding tasks. For example, a user suggested that Copilot could ``\textit{add mapping generator for classes}'' to facilitate the generation of mapping methods between different classes (Discussion \#7870).
    \item \textit{Obsoleted Copilot Version} (OCV) refers to an outdated version of Copilot that is no longer operational. For example, a user encountering an issue with Copilot not working and was informed that ``\textit{your Copilot extension was outdated}'' (Discussion \#17463).
    \item \textit{License Restriction} (LR) refers to situations where some development platforms are unable to be integrated with Copilot due to the restriction imposed by Copilot license. For example, a repository contributor explained that Copilot ``\textit{cannot be used in free or open-source software such as code-server}'' due to its license restriction (Issue \#122).
    \item \textit{Code Telemetry Issue} (CTI) is the reason that some users request new features, aiming to prevent the exposure of their code to Copilot. For example, a user wanted to ``\textit{disable Copilot per workspace}'', so that he could ``\textit{use it for open-source projects but not private/work projects}'' (Discussion \#47991).
\end{itemize}

\begin{table*}[h]
\caption{Mapping between problem types (vertical) and cause types (horizontal)}
\label{tab:issues-causes}
\begin{adjustbox}{width=\textwidth,center}
\begin{tabular}{clcccccccccccccccc}
                                  & & \multicolumn{1}{l}{\textbf{CIE}}     & \multicolumn{1}{l}{\textbf{NCE}} & \multicolumn{1}{l}{\textbf{EICI}} & \multicolumn{1}{l}{\textbf{UP}} & \multicolumn{1}{l}{\textbf{ICS}} & \multicolumn{1}{l}{\textbf{PFE}}& \multicolumn{1}{l}{\textbf{FCH}} & \multicolumn{1}{l}{\textbf{UU}} & 
                                  \multicolumn{1}{l}{\textbf{IUO}} & %\multicolumn{1}{l}{\textbf{FCI}} & 
                                  \multicolumn{1}{l}{\textbf{IDC}} & \multicolumn{1}{l}{\textbf{UF}} & \multicolumn{1}{l}{\textbf{PCI}} & \multicolumn{1}{l}{\textbf{FHCE}} & \multicolumn{1}{l}{\textbf{LR}} & \multicolumn{1}{l}{\textbf{OCV}} & \multicolumn{1}{l}{\textbf{CTI}} \\
\multirow{4}{*}{\textbf{{\normalsize Compatibility Issue}}} &\cellcolor[HTML]{E0F7E0}Editor/IDE Compatibility Issue       & \cellcolor[HTML]{D9E2F3}2 & \cellcolor[HTML]{F2F2F2}0        & \cellcolor[HTML]{D9E2F3}1          & \cellcolor[HTML]{D9E2F3}3          & \cellcolor[HTML]{D9E2F3}3 &\cellcolor[HTML]{F2F2F2}0 &\cellcolor[HTML]{F2F2F2}0     & 
\cellcolor[HTML]{F2F2F2}0          & \cellcolor[HTML]{D9E2F3}1               & \cellcolor[HTML]{D9E2F3}1     & \cellcolor[HTML]{F2F2F2}0          & \cellcolor[HTML]{F2F2F2}0 & \cellcolor[HTML]{F2F2F2}0   & \cellcolor[HTML]{F2F2F2}0          & \cellcolor[HTML]{F2F2F2}0  & \cellcolor[HTML]{F2F2F2}0  \\
% &\cellcolor[HTML]{E0F7E0}Framework Compatibility Issue        & \cellcolor[HTML]{D9E2F3}1 & \cellcolor[HTML]{F2F2F2}0        & \cellcolor[HTML]{F2F2F2}0          & \cellcolor[HTML]{F2F2F2}0          & \cellcolor[HTML]{F2F2F2}0     &\cellcolor[HTML]{F2F2F2}0 &\cellcolor[HTML]{F2F2F2}0 & 
% \cellcolor[HTML]{F2F2F2}0          & \cellcolor[HTML]{F2F2F2}0     & \cellcolor[HTML]{F2F2F2}0          & \cellcolor[HTML]{F2F2F2}0     & \cellcolor[HTML]{F2F2F2}0          & \cellcolor[HTML]{F2F2F2}0  & \cellcolor[HTML]{F2F2F2}0  & \cellcolor[HTML]{F2F2F2}0          & \cellcolor[HTML]{F2F2F2}0  & \cellcolor[HTML]{F2F2F2}0 \\
&\cellcolor[HTML]{E0F7E0}Keyboard Compatibility Issue            & \cellcolor[HTML]{F2F2F2}0 & \cellcolor[HTML]{F2F2F2}0        & \cellcolor[HTML]{F2F2F2}0          & \cellcolor[HTML]{F2F2F2}0          & \cellcolor[HTML]{F2F2F2}0     &\cellcolor[HTML]{F2F2F2}0 &\cellcolor[HTML]{F2F2F2}0 & 
\cellcolor[HTML]{F2F2F2}0       & \cellcolor[HTML]{F2F2F2}0          & \cellcolor[HTML]{F2F2F2}0     & \cellcolor[HTML]{F2F2F2}0          & \cellcolor[HTML]{F2F2F2}0  & \cellcolor[HTML]{F2F2F2}0  & \cellcolor[HTML]{F2F2F2}0          & \cellcolor[HTML]{F2F2F2}0 & \cellcolor[HTML]{F2F2F2}0  \\
&\cellcolor[HTML]{E0F7E0}Plug-in Compatibility Issue       & \cellcolor[HTML]{D9E2F3}2 & \cellcolor[HTML]{F2F2F2}0        & \cellcolor[HTML]{F2F2F2}0          & \cellcolor[HTML]{F2F2F2}0          & \cellcolor[HTML]{D9E2F3}4    &\cellcolor[HTML]{F2F2F2}0 &\cellcolor[HTML]{F2F2F2}0  & 
\cellcolor[HTML]{F2F2F2}0          & \cellcolor[HTML]{F2F2F2}0         & \cellcolor[HTML]{D9E2F3}3     & \cellcolor[HTML]{F2F2F2}0          & \cellcolor[HTML]{F2F2F2}0 & \cellcolor[HTML]{F2F2F2}0   & \cellcolor[HTML]{F2F2F2}0          & \cellcolor[HTML]{F2F2F2}0 & \cellcolor[HTML]{F2F2F2}0  \\
\multirow{3}{*}{\textbf{{\normalsize Copyright and Policy Issue}}} &\cellcolor[HTML]{FFFFE0}Code Telemetry Issue                     & \cellcolor[HTML]{F2F2F2}0 & \cellcolor[HTML]{F2F2F2}0        & \cellcolor[HTML]{F2F2F2}0          & \cellcolor[HTML]{F2F2F2}0    &\cellcolor[HTML]{F2F2F2}0 &\cellcolor[HTML]{F2F2F2}0  & 
\cellcolor[HTML]{F2F2F2}0          & \cellcolor[HTML]{F2F2F2}0     & \cellcolor[HTML]{F2F2F2}0          & \cellcolor[HTML]{F2F2F2}0     & \cellcolor[HTML]{F2F2F2}0          & \cellcolor[HTML]{F2F2F2}0 & \cellcolor[HTML]{F2F2F2}0   & \cellcolor[HTML]{F2F2F2}0          & \cellcolor[HTML]{F2F2F2}0  & \cellcolor[HTML]{F2F2F2}0 \\
& \cellcolor[HTML]{FFFFE0}Code Copyright Issue                    & \cellcolor[HTML]{F2F2F2}0 & \cellcolor[HTML]{F2F2F2}0        & \cellcolor[HTML]{F2F2F2}0          & \cellcolor[HTML]{F2F2F2}0     &\cellcolor[HTML]{F2F2F2}0 &\cellcolor[HTML]{F2F2F2}0 & 
\cellcolor[HTML]{F2F2F2}0          & \cellcolor[HTML]{F2F2F2}0     & \cellcolor[HTML]{F2F2F2}0          & \cellcolor[HTML]{F2F2F2}0     & \cellcolor[HTML]{F2F2F2}0          & \cellcolor[HTML]{F2F2F2}0  & \cellcolor[HTML]{F2F2F2}0  & \cellcolor[HTML]{F2F2F2}0          & \cellcolor[HTML]{F2F2F2}0 & \cellcolor[HTML]{F2F2F2}0  \\
& \cellcolor[HTML]{FFFFE0}Violation of Marketplace Policy   & \cellcolor[HTML]{F2F2F2}0       & \cellcolor[HTML]{F2F2F2}0          & \cellcolor[HTML]{F2F2F2}0          & \cellcolor[HTML]{F2F2F2}0    &\cellcolor[HTML]{F2F2F2}0 &\cellcolor[HTML]{F2F2F2}0  & 
\cellcolor[HTML]{F2F2F2}0          & \cellcolor[HTML]{F2F2F2}0     & \cellcolor[HTML]{F2F2F2}0          & \cellcolor[HTML]{F2F2F2}0     & \cellcolor[HTML]{F2F2F2}0          & \cellcolor[HTML]{F2F2F2}0  & \cellcolor[HTML]{F2F2F2}0  & \cellcolor[HTML]{F2F2F2}0          & \cellcolor[HTML]{F2F2F2}0 & \cellcolor[HTML]{F2F2F2}0  \\
\multirow{4}{*}{\textbf{{\normalsize Feature Request}}} &\cellcolor[HTML]{E0F7E0}Function Request                  & \cellcolor[HTML]{F2F2F2}0 & \cellcolor[HTML]{F2F2F2}0        & \cellcolor[HTML]{F2F2F2}0          & \cellcolor[HTML]{F2F2F2}0          & \cellcolor[HTML]{F2F2F2}0     &\cellcolor[HTML]{4472C4}21 &\cellcolor[HTML]{8EAADB}11 & 
\cellcolor[HTML]{F2F2F2}0          & \cellcolor[HTML]{F2F2F2}0     & \cellcolor[HTML]{F2F2F2}0          & \cellcolor[HTML]{F2F2F2}0     & \cellcolor[HTML]{F2F2F2}0  & \cellcolor[HTML]{D9E2F3}8   & \cellcolor[HTML]{F2F2F2}0          & \cellcolor[HTML]{F2F2F2}0  & \cellcolor[HTML]{D9E2F3}4 \\
& \cellcolor[HTML]{E0F7E0}Integration Request         & \cellcolor[HTML]{F2F2F2}0 & \cellcolor[HTML]{F2F2F2}0        & \cellcolor[HTML]{F2F2F2}0          & \cellcolor[HTML]{F2F2F2}0          & \cellcolor[HTML]{F2F2F2}0    &\cellcolor[HTML]{F2F2F2}0 &\cellcolor[HTML]{8EAADB}14  & 
\cellcolor[HTML]{F2F2F2}0          & \cellcolor[HTML]{F2F2F2}0     & \cellcolor[HTML]{F2F2F2}0          & \cellcolor[HTML]{F2F2F2}0     & \cellcolor[HTML]{F2F2F2}0          & \cellcolor[HTML]{D9E2F3}1   & \cellcolor[HTML]{F2F2F2}0          & \cellcolor[HTML]{F2F2F2}0  & \cellcolor[HTML]{F2F2F2}0 \\
& \cellcolor[HTML]{E0F7E0}Professional Copilot Version             & \cellcolor[HTML]{F2F2F2}0        & \cellcolor[HTML]{F2F2F2}0          & \cellcolor[HTML]{F2F2F2}0          & \cellcolor[HTML]{F2F2F2}0    &\cellcolor[HTML]{F2F2F2}0 &\cellcolor[HTML]{F2F2F2}0  & 
\cellcolor[HTML]{F2F2F2}0          & \cellcolor[HTML]{F2F2F2}0     & \cellcolor[HTML]{F2F2F2}0          & \cellcolor[HTML]{F2F2F2}0     & \cellcolor[HTML]{F2F2F2}0          & \cellcolor[HTML]{F2F2F2}0 & \cellcolor[HTML]{F2F2F2}0   & \cellcolor[HTML]{F2F2F2}0          & \cellcolor[HTML]{F2F2F2}0  & \cellcolor[HTML]{F2F2F2}0 \\
& \cellcolor[HTML]{E0F7E0}UI Request                     & \cellcolor[HTML]{F2F2F2}0 & \cellcolor[HTML]{F2F2F2}0        & \cellcolor[HTML]{F2F2F2}0          & \cellcolor[HTML]{F2F2F2}0          & \cellcolor[HTML]{F2F2F2}0    &\cellcolor[HTML]{D9E2F3}5   & 
\cellcolor[HTML]{F2F2F2}0          & \cellcolor[HTML]{F2F2F2}0     & \cellcolor[HTML]{F2F2F2}0          & \cellcolor[HTML]{F2F2F2}0     & \cellcolor[HTML]{F2F2F2}0          & \cellcolor[HTML]{F2F2F2}0 & \cellcolor[HTML]{F2F2F2}0   & \cellcolor[HTML]{F2F2F2}0          & \cellcolor[HTML]{F2F2F2}0 & \cellcolor[HTML]{F2F2F2}0  \\
\multirow{7}{*}{\textbf{{\normalsize Suggestion Content Issue}}} &\cellcolor[HTML]{FFFFE0}Insecure Suggestion                 & \cellcolor[HTML]{F2F2F2}0        & \cellcolor[HTML]{F2F2F2}0          & \cellcolor[HTML]{F2F2F2}0          & \cellcolor[HTML]{F2F2F2}0    &\cellcolor[HTML]{F2F2F2}0 &\cellcolor[HTML]{F2F2F2}0  & 
\cellcolor[HTML]{F2F2F2}0          & \cellcolor[HTML]{F2F2F2}0     & \cellcolor[HTML]{F2F2F2}0          & \cellcolor[HTML]{F2F2F2}0     & \cellcolor[HTML]{F2F2F2}0          & \cellcolor[HTML]{F2F2F2}0  & \cellcolor[HTML]{F2F2F2}0  & \cellcolor[HTML]{F2F2F2}0          & \cellcolor[HTML]{F2F2F2}0  & \cellcolor[HTML]{F2F2F2}0 \\
& \cellcolor[HTML]{FFFFE0}Less Efficient Suggestion                  & \cellcolor[HTML]{F2F2F2}0 & \cellcolor[HTML]{F2F2F2}0        & \cellcolor[HTML]{F2F2F2}0               & \cellcolor[HTML]{F2F2F2}0    &\cellcolor[HTML]{F2F2F2}0 &\cellcolor[HTML]{F2F2F2}0  & 
\cellcolor[HTML]{F2F2F2}0          & \cellcolor[HTML]{F2F2F2}0     & \cellcolor[HTML]{F2F2F2}0          & \cellcolor[HTML]{F2F2F2}0     & \cellcolor[HTML]{F2F2F2}0          & \cellcolor[HTML]{F2F2F2}0  & \cellcolor[HTML]{F2F2F2}0  & \cellcolor[HTML]{F2F2F2}0          & \cellcolor[HTML]{F2F2F2}0  & \cellcolor[HTML]{F2F2F2}0 \\
& \cellcolor[HTML]{FFFFE0}Low Quality Suggestion                 & \cellcolor[HTML]{D9E2F3}1         & \cellcolor[HTML]{F2F2F2}0          & \cellcolor[HTML]{F2F2F2}0          & \cellcolor[HTML]{F2F2F2}0     &\cellcolor[HTML]{F2F2F2}0 &\cellcolor[HTML]{F2F2F2}0 & 
\cellcolor[HTML]{F2F2F2}0          & \cellcolor[HTML]{F2F2F2}0     & \cellcolor[HTML]{F2F2F2}0          & \cellcolor[HTML]{F2F2F2}0     & \cellcolor[HTML]{D9E2F3}2          & \cellcolor[HTML]{F2F2F2}0 & \cellcolor[HTML]{F2F2F2}0   & \cellcolor[HTML]{F2F2F2}0          & \cellcolor[HTML]{F2F2F2}0  & \cellcolor[HTML]{F2F2F2}0 \\
& \cellcolor[HTML]{FFFFE0}Nonsensical Suggestion                 & \cellcolor[HTML]{D9E2F3}1 & \cellcolor[HTML]{F2F2F2}0        & \cellcolor[HTML]{F2F2F2}0          & \cellcolor[HTML]{F2F2F2}0          & \cellcolor[HTML]{D9E2F3}1      &\cellcolor[HTML]{F2F2F2}0 & 
\cellcolor[HTML]{F2F2F2}0          & \cellcolor[HTML]{F2F2F2}0     & \cellcolor[HTML]{F2F2F2}0          & \cellcolor[HTML]{F2F2F2}0     & \cellcolor[HTML]{F2F2F2}0          & \cellcolor[HTML]{F2F2F2}0 & \cellcolor[HTML]{F2F2F2}0   & \cellcolor[HTML]{F2F2F2}0          & \cellcolor[HTML]{F2F2F2}0  & \cellcolor[HTML]{F2F2F2}0 \\
& \cellcolor[HTML]{FFFFE0}Suggestion with Bugs                & \cellcolor[HTML]{F2F2F2}0        & \cellcolor[HTML]{F2F2F2}0          & \cellcolor[HTML]{F2F2F2}0          & \cellcolor[HTML]{F2F2F2}0    &\cellcolor[HTML]{F2F2F2}0 &\cellcolor[HTML]{F2F2F2}0  & 
\cellcolor[HTML]{F2F2F2}0          & \cellcolor[HTML]{F2F2F2}0     & \cellcolor[HTML]{F2F2F2}0          & \cellcolor[HTML]{F2F2F2}0     & \cellcolor[HTML]{F2F2F2}0          & \cellcolor[HTML]{F2F2F2}0 & \cellcolor[HTML]{F2F2F2}0   & \cellcolor[HTML]{F2F2F2}0          & \cellcolor[HTML]{F2F2F2}0 & \cellcolor[HTML]{F2F2F2}0  \\
& \cellcolor[HTML]{FFFFE0}Incomprehensible Suggestion         & \cellcolor[HTML]{F2F2F2}0         & \cellcolor[HTML]{F2F2F2}0          & \cellcolor[HTML]{F2F2F2}0          & \cellcolor[HTML]{F2F2F2}0     &\cellcolor[HTML]{F2F2F2}0 &\cellcolor[HTML]{F2F2F2}0 & 
\cellcolor[HTML]{F2F2F2}0          & \cellcolor[HTML]{F2F2F2}0     & \cellcolor[HTML]{F2F2F2}0          & \cellcolor[HTML]{F2F2F2}0     & \cellcolor[HTML]{F2F2F2}0          & \cellcolor[HTML]{F2F2F2}0 & \cellcolor[HTML]{F2F2F2}0   & \cellcolor[HTML]{F2F2F2}0          & \cellcolor[HTML]{F2F2F2}0 & \cellcolor[HTML]{F2F2F2}0  \\
& \cellcolor[HTML]{FFFFE0}Suggestion with Invalid Syntax          & \cellcolor[HTML]{F2F2F2}0         & \cellcolor[HTML]{F2F2F2}0          & \cellcolor[HTML]{F2F2F2}0          & \cellcolor[HTML]{F2F2F2}0     &\cellcolor[HTML]{F2F2F2}0 &\cellcolor[HTML]{F2F2F2}0 & 
\cellcolor[HTML]{F2F2F2}0          & \cellcolor[HTML]{F2F2F2}0     & \cellcolor[HTML]{F2F2F2}0          & \cellcolor[HTML]{F2F2F2}0     & \cellcolor[HTML]{F2F2F2}0          & \cellcolor[HTML]{F2F2F2}0  & \cellcolor[HTML]{F2F2F2}0  & \cellcolor[HTML]{F2F2F2}0          & \cellcolor[HTML]{F2F2F2}0  & \cellcolor[HTML]{F2F2F2}0 \\
\multirow{6}{*}{\textbf{{\normalsize Operation Issue}}} & \cellcolor[HTML]{E0F7E0}Accessing Failure          & \cellcolor[HTML]{D9E2F3}3 & \cellcolor[HTML]{8EAADB}11        & \cellcolor[HTML]{F2F2F2}0          & \cellcolor[HTML]{D9E2F3}1          & \cellcolor[HTML]{F2F2F2}0     &\cellcolor[HTML]{F2F2F2}0 &\cellcolor[HTML]{F2F2F2}0 & 
\cellcolor[HTML]{D9E2F3}4          & \cellcolor[HTML]{F2F2F2}0          & \cellcolor[HTML]{D9E2F3}1     & \cellcolor[HTML]{D9E2F3}3          & \cellcolor[HTML]{F2F2F2}0  & \cellcolor[HTML]{F2F2F2}0  & \cellcolor[HTML]{F2F2F2}0          & \cellcolor[HTML]{F2F2F2}0  & \cellcolor[HTML]{F2F2F2}0 \\
& \cellcolor[HTML]{E0F7E0}Authentication Failure          & \cellcolor[HTML]{4472C4}25 & \cellcolor[HTML]{8EAADB}16        & \cellcolor[HTML]{D9E2F3}8          & \cellcolor[HTML]{D9E2F3}9          & \cellcolor[HTML]{D9E2F3}2    &\cellcolor[HTML]{F2F2F2}0 &\cellcolor[HTML]{F2F2F2}0  & 
\cellcolor[HTML]{8EAADB}13          & \cellcolor[HTML]{D9E2F3}5              & \cellcolor[HTML]{F2F2F2}0     & \cellcolor[HTML]{D9E2F3}2          & \cellcolor[HTML]{F2F2F2}0 & \cellcolor[HTML]{F2F2F2}0   & \cellcolor[HTML]{D9E2F3}1          & \cellcolor[HTML]{D9E2F3}1 & \cellcolor[HTML]{F2F2F2}0  \\
& \cellcolor[HTML]{E0F7E0}Functionality Failure           & \cellcolor[HTML]{8EAADB}17 & \cellcolor[HTML]{D9E2F3}6        & \cellcolor[HTML]{8EAADB}10          & \cellcolor[HTML]{F2F2F2}0          & \cellcolor[HTML]{8EAADB}14    &\cellcolor[HTML]{F2F2F2}0 &\cellcolor[HTML]{F2F2F2}0  & 
\cellcolor[HTML]{D9E2F3}1          & \cellcolor[HTML]{D9E2F3}4          & \cellcolor[HTML]{D9E2F3}6     & \cellcolor[HTML]{D9E2F3}5          & \cellcolor[HTML]{8EAADB}11 & \cellcolor[HTML]{F2F2F2}0   & \cellcolor[HTML]{F2F2F2}0          & \cellcolor[HTML]{D9E2F3}3  & \cellcolor[HTML]{F2F2F2}0 \\
& \cellcolor[HTML]{E0F7E0}Installation Issue                & \cellcolor[HTML]{F2F2F2}0 & \cellcolor[HTML]{F2F2F2}0        & \cellcolor[HTML]{8EAADB}11          & \cellcolor[HTML]{8EAADB}12            &\cellcolor[HTML]{F2F2F2}0 &\cellcolor[HTML]{F2F2F2}0 & 
\cellcolor[HTML]{F2F2F2}0          & \cellcolor[HTML]{F2F2F2}0     & \cellcolor[HTML]{F2F2F2}0          & \cellcolor[HTML]{F2F2F2}0     & \cellcolor[HTML]{F2F2F2}0          & \cellcolor[HTML]{F2F2F2}0 & \cellcolor[HTML]{F2F2F2}0   & \cellcolor[HTML]{D9E2F3}2          & \cellcolor[HTML]{F2F2F2}0  & \cellcolor[HTML]{F2F2F2}0 \\
& \cellcolor[HTML]{E0F7E0}Startup Issue          & \cellcolor[HTML]{8EAADB}19 & \cellcolor[HTML]{8EAADB}19        & \cellcolor[HTML]{8EAADB}15          & \cellcolor[HTML]{D9E2F3}4          & \cellcolor[HTML]{D9E2F3}3    &\cellcolor[HTML]{F2F2F2}0 &\cellcolor[HTML]{F2F2F2}0  & 
\cellcolor[HTML]{D9E2F3}4          & \cellcolor[HTML]{D9E2F3}5              & \cellcolor[HTML]{F2F2F2}0     & \cellcolor[HTML]{F2F2F2}0          & \cellcolor[HTML]{D9E2F3}3 & \cellcolor[HTML]{F2F2F2}0   & \cellcolor[HTML]{D9E2F3}1          & \cellcolor[HTML]{D9E2F3}1 & \cellcolor[HTML]{F2F2F2}0  \\
& \cellcolor[HTML]{E0F7E0}Version Control Issue          & \cellcolor[HTML]{D9E2F3}2 & \cellcolor[HTML]{F2F2F2}0        & \cellcolor[HTML]{D9E2F3}4          & \cellcolor[HTML]{D9E2F3}3          & \cellcolor[HTML]{F2F2F2}0    &\cellcolor[HTML]{F2F2F2}0 &\cellcolor[HTML]{F2F2F2}0           & \cellcolor[HTML]{F2F2F2}0     & \cellcolor[HTML]{F2F2F2}0          & \cellcolor[HTML]{F2F2F2}0     & \cellcolor[HTML]{F2F2F2}0          & \cellcolor[HTML]{F2F2F2}0  & \cellcolor[HTML]{F2F2F2}0  & \cellcolor[HTML]{F2F2F2}0          & \cellcolor[HTML]{D9E2F3}1 & \cellcolor[HTML]{F2F2F2}0  \\
\multirow{4}{*}{\textbf{{\normalsize User Experience Issue}}} &\cellcolor[HTML]{FFFFE0}Poor Authentication Experience           & \cellcolor[HTML]{D9E2F3}2 & \cellcolor[HTML]{F2F2F2}0        & \cellcolor[HTML]{D9E2F3}1                   & \cellcolor[HTML]{F2F2F2}0    &\cellcolor[HTML]{F2F2F2}0 &\cellcolor[HTML]{F2F2F2}0  & 
\cellcolor[HTML]{F2F2F2}0          & \cellcolor[HTML]{F2F2F2}0     & \cellcolor[HTML]{F2F2F2}0          & \cellcolor[HTML]{F2F2F2}0     & \cellcolor[HTML]{F2F2F2}0          & \cellcolor[HTML]{F2F2F2}0 & \cellcolor[HTML]{F2F2F2}0   & \cellcolor[HTML]{F2F2F2}0          & \cellcolor[HTML]{D9E2F3}1  & \cellcolor[HTML]{F2F2F2}0 \\
& \cellcolor[HTML]{FFFFE0}Poor Functionality Experience           & \cellcolor[HTML]{F2F2F2}0 & \cellcolor[HTML]{F2F2F2}0        & \cellcolor[HTML]{F2F2F2}0          & \cellcolor[HTML]{F2F2F2}0          & \cellcolor[HTML]{F2F2F2}0    &\cellcolor[HTML]{F2F2F2}0 &\cellcolor[HTML]{F2F2F2}0  & 
\cellcolor[HTML]{F2F2F2}0          & \cellcolor[HTML]{F2F2F2}0               & \cellcolor[HTML]{D9E2F3}1     & \cellcolor[HTML]{F2F2F2}0          & \cellcolor[HTML]{F2F2F2}0 & \cellcolor[HTML]{F2F2F2}0   & \cellcolor[HTML]{F2F2F2}0          & \cellcolor[HTML]{F2F2F2}0 & \cellcolor[HTML]{F2F2F2}0  \\
& \cellcolor[HTML]{FFFFE0}Poor Performance                  & \cellcolor[HTML]{D9E2F3}2 & \cellcolor[HTML]{F2F2F2}0        & \cellcolor[HTML]{F2F2F2}0          & \cellcolor[HTML]{F2F2F2}0          & \cellcolor[HTML]{F2F2F2}0    &\cellcolor[HTML]{F2F2F2}0 &\cellcolor[HTML]{F2F2F2}0  & 
\cellcolor[HTML]{F2F2F2}0          & \cellcolor[HTML]{D9E2F3}1               & \cellcolor[HTML]{F2F2F2}0     & \cellcolor[HTML]{F2F2F2}0          & \cellcolor[HTML]{D9E2F3}1 & \cellcolor[HTML]{F2F2F2}0   & \cellcolor[HTML]{F2F2F2}0          & \cellcolor[HTML]{F2F2F2}0  & \cellcolor[HTML]{F2F2F2}0 \\
& \cellcolor[HTML]{FFFFE0}Poor Subscription Experience            & \cellcolor[HTML]{F2F2F2}0 & \cellcolor[HTML]{D9E2F3}1        & \cellcolor[HTML]{F2F2F2}0          & \cellcolor[HTML]{F2F2F2}0          & \cellcolor[HTML]{F2F2F2}0    &\cellcolor[HTML]{F2F2F2}0 &\cellcolor[HTML]{F2F2F2}0  & 
\cellcolor[HTML]{F2F2F2}0          & \cellcolor[HTML]{D9E2F3}1            & \cellcolor[HTML]{F2F2F2}0     & \cellcolor[HTML]{F2F2F2}0          & \cellcolor[HTML]{F2F2F2}0 & \cellcolor[HTML]{F2F2F2}0   & \cellcolor[HTML]{F2F2F2}0          & \cellcolor[HTML]{F2F2F2}0  & \cellcolor[HTML]{F2F2F2}0 \\
\end{tabular}
\end{adjustbox}
\begin{minipage}{18cm} 
\vspace{0.1cm}
\vspace{0.1cm}
\scriptsize  \textbf{Full names of each cause type:} CIE: \textit{Copilot Internal Error}; NCE: \textit{Network Connection Error}; EICI: \textit{Editor/IDE Compatibility Issue}; UP: \textit{Unsupported Platform}; ICS:  \textit{Improper Configuration/Setting}; PFE: \textit{Poor Functionality Experience}; FCH: \textit{For Coding Habit}; UU: \textit{User Unauthorized}; IUO: \textit{Improper User Operation}; IDC: \textit{Intentional Design of Copilot}; UF: \textit{Unimplemented Feature}; PCI: \textit{Plug-in Compatibility Issue}; FHCE: \textit{For Higher Coding Efficiency}; LR: \textit{License Restriction}; OCV: \textit{Obsoleted Copilot Version}; CTI: \textit{Code Telemetry Issue}.
\end{minipage}
\end{table*}

\subsubsection{Causes to Problems Mapping}
Table~\ref{tab:issues-causes} illustrates the mapping relationship of Copilot related problems to their causes. We use abbreviations to represent each type of cause; for example, ``CIE'' represents \textit{Copilot Internal Error}. 
% Table \ref{The Types of causes leading to the issues when using Copilot} presents the full names for the abbreviations of all the cause types.
The full names for all types of causes are provided in the note of Table \ref{tab:issues-causes}.

Over one third of \textit{Operation Issues} (37.2\%) have associated causes. Specifically, \textsc{authentication failure} is primarily induced by CIE, NCE, and UU; \textsc{functionality failure} is mainly caused by CIE, ICS, and PCI; The occurrence of \textsc{installation issue} typically stems from EICI and UP; \textsc{startup issue} commonly originates from CIE and NCE; while \textsc{accessing failure} is primarily attributed to NCE; and \textsc{version control issue} is mainly brought about by CIE, EICI, and UP.
% The causes of \textsc{version control issue} were relatively fewer than other types of problems in \textit{Operation Issue}.

For \textit{Compatibility Issue}, causes are identified in 9.5\% of the cases. To be specific, \textsc{editor/ide compatibility issue} is mainly caused by UP and ICS, while \textsc{plug-in compatibility issue} is attributed to ICS and IDC.

For \textit{Feature Request}, four types of causes are identified in 31.0\% of cases, which are FCH, PFE, FHCE, and CTI. PFE and FCH are the prime causes for users to raise \textsc{function requests}, while \textsc{integration requests} are mainly attributed to FCH.

For \textit{User Experience Issue}, causes are identified in 19.0\% of the cases. However, there are only a few causes identified for each type of \textit{User Experience Issue}, with CIE identified as the prime cause for \textsc{poor authentication experience} and \textsc{poor performance}.
% \textcolor{blue}{For \textit{Suggestion Content Issue}, only a limited number of causes have been identified, which are exclusively attributed to two types: \textsc{low quality suggestion} and \textsc{nonsensical suggestion}. The specific causes are CII, ICS, and UF.}

%For \textit{Suggestion Content Issue}, only a few cases of \textsc{low quality suggestions} and \textsc{nonsensical suggestions} have had their causes identified, which are CIE, ICS, and UF.

For \textit{Suggestion Content Issue}, causes are identified in 8.6\% of the cases. CIE and UF are the causes leading to \textsc{low quality suggestion}, while CIE and ICS are the causes for \textsc{nonsensical suggestion}.

\textit{Copyright and Policy Issue} refers to the concerns of some users about code leakage to Copilot, which inherently forms the reason for raising such problems, thereby no need for further identification of the underlying causes.

\subsubsection{\textbf{Interpretation}}
\textbf{Frequency of Causes}: CIE is the most common type of cause leading to Copilot usage problems. Typically, the cause identification of CIE relies on user feedback regarding abnormal usage experiences of Copilot, and it often results in a group of users reporting the same problem within a certain time period. For example, ``\textit{a bad deployment}'' of the Copilot server caused a group of users to report \textsc{authentication failures} (Discussion \#39533). The high number of NCE, EICI, and ICS indicates that some problems arise from the environment in which Copilot operates. A common situation of NCE is that users connect to the Copilot server through an HTTP proxy, which may lead to the intercept of Secure Socket Layer (SSL). However, Copilot now offers support for access through an HTTP proxy, thus addressing such problems~\citep{copilotproxysupport}. PFE, FCH, FCE, and CTI are the four types of causes for \textit{Feature Request}. The remaining eight types of causes are less common, but can still provide insights into specific problems related to Copilot. For instance, UU is identified as the direct cause for many users experiencing \textsc{authentication failure} and \textsc{functionality failure} when using Copilot.

\textbf{Mapping of Causes to Problems}:
% In \textit{Operation Issue}, a variety of causes have been identified, reflective of the complexity and diversity of these problems, which typically take precedence for user resolution. 
When Copilot users encounter \textit{Operation Issues}, nearly one quarter (23.1\%) of these problems are caused by errors originating from Copilot server (i.e. CIE), while the rest of such problems are induced by the environment on which Copilot operates. 
According to the causes of \textit{Feature Request}, it appears that users typically request for new functionalities or enhancement of existing features based on their personal coding habits and suboptimal experiences when coding with Copilot. 
Fewer causes have been identified for \textit{Compatibility Issue}. When users identify the problem as \textit{Compatibility Issue}, they tend to focus on finding solutions rather than further analyzing the causes. 
The causes identified for \textit{Suggestion Content Issue} and \textit{Copyright and Policy Issue} are limited in number. One possible reason is that the non-open source nature of Copilot prevents users from investigating the causes of the problems in these two categories.
%The causes leading to \textit{User Experience Issue} do not exhibit a clear pattern in their distribution. 
From the results of Table \ref{tab:issues-causes}, we did not identify the main causes leading to \textit{User Experience Issue}. However, when users encounter \textsc{poor authentication experience} or \textsc{poor performance} when using Copilot, consideration should be given to whether there are internal errors within Copilot, as the four cases of CIE leading to such problems.

\subsection{Type of Solutions (RQ3)}
\label{subsec:solutionsresults}
%\noindent \textbf{RQ3: What are the potential solutions to address these issues?}\\
%We collected and analyzed solutions from both the Copilot team and users that have been confirmed to effectively solve Copilot related issues, which are presented in Table \ref{The types of solutions to the issues when using Copilot}. 

\subsubsection{\textbf{Results}}
As mentioned in Section \ref{sec:formalDataExtraction}, not all problems have corresponding solutions that can be extracted. As a result, we identified a total of 497 solutions, which were used to address 36.7\% of all problems, and categorized into 11 types as shown in Table \ref{The types of solutions to the issues when using Copilot}. The result reveals that most of the usage bugs were addressed by \textit{Bug fixed by Copilot} (27.2\%). When users tried to solve problems themselves, \textit{Modify Configuration/Setting} (22.1\%), \textit{Use Suitable Version} (17.1\%), and \textit{Reinstall/Restart/Reauthorize Copilot} (12.1\%) were commonly used as effective solutions. The example, count, and proportion of each type of solution are presented in Table \ref{The types of solutions to the issues when using Copilot}. 
%Due to space constraints, we will focus on interpreting the results of the top 5 solutions. It should be noted that \textit{Others} (4.2\%) is a collection of dedicated solutions that are usually specific to particular environments and problems.

\begin{itemize}
    \item \textit{Bug Fixed by Copilot} (BFC) refers to the process where the Copilot team collects problems reported by users or conducts functionality test on Copilot, and then addresses identified bugs. For example, a user failed to use Copilot because of the ``\textit{problem from Copilot's server-side}'', and this problem was fixed by the Copilot team as soon as they noticed it (Issue \#86). Typically, the resolution process to the problems by the Copilot team is unclear for users.
    \item \textit{Modify Configuration/Setting} (MCS) refers to the solutions that users adjust settings or configuration files of Copilot to solve specific usage problems. 
    %Taking the use of Copilot in VSCode as an example, modifying files such as 'keybindings.json' and 'settings.json', can enable users to troubleshoot and resolve some issues with Copilot on their own. 
    For example, by ``\textit{updating the file setting.json}'', a user resolved the problem of Copilot ``\textit{not adding closing brackets}'' (SO \#68347605).
    \item \textit{Use Suitable Version} (USV) refers to users choosing a suitable version of Copilot that resolves the problems faced in the Copilot version currently used. This type of solution also includes adjusting the version of IDEs or code editors to be compatible with Copilot. For example, a user who wanted to use Copilot in Visual Studio solved the problem of ``\textit{not being able to find the GitHub copilot extension}'', by ``\textit{upgrading Visual Studio to 17.2}'' (Discussion \#18566).
    \item \textit{Reinstall/Restart/Reauthorize Copilot} (RC) is a type of solution allowing users to reset Copilot to its original state, thereby resolving any previous errors or restoring settings to their default conditions. For example, a user found that ``\textit{GitHub Copilot could not connect to server}'' in VSCode, and solved it by simply ``\textit{restarting extension}'' (Discussion \#27378).
    \item \textit{Feature Implemented by Copilot} (FIC) refers to the introduction or enhancement of a feature by the Copilot team. For example, the latest Copilot ``\textit{contains preliminary support for connecting through HTTP proxy servers}'', addressing the \textsc{authentication failure} previously experienced by some users (SO \#73242748).
    \item \textit{Follow Official Instruction} (FOI) refers to the process where users follow the steps provided by the Copilot UI or described in the user manuals during registration, login, subscription, and configuration. This type of solution usually aimed at assisting users who are not familiar with Copilot. For example, a user was advised to consult the ``\textit{Copilot doc}'' for guidance on configuring keyboard shortcuts on Mac (SO \#75276040).
    \item \textit{Disable Interfering Factors} (DIF) refers to the process where users identify and remove the factors disrupting the normal operation of Copilot, such as other activated plug-ins and HTTP proxies that could interfere with network connection. For example, a user reported that ``\textit{GitHub Copilot autosuggesions are not auto-filled in .md files}'', and solved this problem by disabling the ``\textit{Markdown all in one}'' plug-in (Discussion \#10203).
    \item \textit{Restart Runtime Environment} (RRE) refers to the process of restarting the environment Copilot operates on to resolve any errors affecting the current runtime environment. For example, restarting the computer and IDE on which Copilot is running helped a user solve an \textsc{installation failure} on IntelliJ IDEA 2022 (Discussion \#17638).
    \item \textit{Modify the Input Way} (MIW) refers to users changing the way they prompt Copilot to generate code, to address problems that Copilot does not automatically provide code suggestions, or to make Copilot provide codes that better meet their expectations. For example, ``\textit{going to a new line after a code comment, and starting typing what you want to create}'' resolved the problem that Copilot created ``\textit{the same comment over and over again}'' (SO \#70995718).
    \item \textit{Install/Update Framework} (IUF) refers to installing or updating the frameworks in IDEs and code editors for Copilot to operate. This solution is primarily employed to resolve the incompatibility between Copilot.vim \citep{copilotvim} and Node.js. For example, a user found that he had to ``\textit{install node.js}'' for Copilot to work in Neovim (Discussion \#40300).
    \item \textit{Others}: Different from Copilot usage problems and their causes, which can be categorized into specific categories and types, several dedicated solutions aimed at addressing certain Copilot usage problems are classified under the category \textit{Others}. For example, using the VSCode extension ``\textit{Win-CA}'' in append mode to resolve \textsc{accessing failure} of Copilot was identified only once, making it difficult to form a new type of solution (SO \#71367058).
    %some solutions elude classification through Constant Comparison and are assigned to the \textit{Others} category. These solutions typically involve an array of detailed operational steps and lack a sufficient number of analogous methods to facilitate comparison. As a result, the distinctiveness of these solutions precludes their recognition as a separate type of solution. For example, using the VSCode extension ``\textit{Win-CA}'' in append mode to resolve \textsc{accessing failure} of Copilot was identified only once, making it difficult to form a new type of solution or to classify it into an existing one (SO \#71367058).
\end{itemize}

\begin{table*}[htbp]
\renewcommand{\arraystretch}{1.4}
\scriptsize
% \small
\caption{Solutions of Copilot usage problems}
\label{The types of solutions to the issues when using Copilot}
    
%\begin{tabular}{m{3.35cm}m{11.9cm}m{0.6cm}<{\centering}m{0.6cm}<{\centering}}
\begin{tabular}{m{5cm}m{9cm}m{0.8cm}<{\centering}m{0.8cm}<{\centering}}
\hline
    \textbf{Type}                   & \textbf{Example (Extracted Solution)}                                                                      & \textbf{Count}   & \textbf{\%}           \\ \hline
% BFC (Bug Fixed by Copilot)         & \textit{Plz update to 1.40.0.1 which contains a fix.} (Discussion \#29693)                                                   & 134               & 27.0\% \\ \hline
Bug Fixed by Copilot (BFC)         & \textit{The problem is from Copilot's server-side and got fixed now.} (Issue \#86)                                                   & 135               & 27.2\% \\ \hline
% MCS (Modify Configuration/Setting)      & \textit{You can change the hotkey at Preferences > Keymap > Plugins > GitHub Copilot.} (Discussion \#15258)        & 110               & 22.1\% \\ \hline
Modify Configuration/Setting (MCS)      & \textit{Update your settings.json like the following...} (SO \#68347605)        & 110               & 22.1\% \\ \hline
Use Suitable Version   (USV)           & \textit{Upgrading to 17.2 solved the problem.} (Discussion \#18566) & 85               & 17.1\% \\ \hline
% RC (Reinstall/Restart/Reauthorize Copilot)           & \textit{Please try signing out, copilot will prompt you to sign back in.} (Discussion \#32867)     & 61               & 12.3\% \\ \hline
Reinstall/Restart/Reauthorize Copilot (RC)          & \textit{Problem solved after restarting extension.} (Discussion \#27378)     & 60               & 12.1\% \\ \hline
% FIC (Feature Implemented by Copilot) & \textit{1.1.3 has been released and contains the new setting.} (Discussion \#8444)   & 32               & 6.4\% \\ \hline
Feature Implemented by Copilot (FIC) & \textit{The latest release of Copilot for Visual Studio Code and Copilot for IntelliJ contains preliminary support for connecting through HTTP proxy servers.} (SO \#73242748)   & 32               & 6.4\% \\ \hline
Follow Official Instruction (FOI) & \textit{In the Copilot docs, see for Visual Studio Code ``\textit{Keyboard shortcuts for GitHub Copilot}'' } (SO \#75276040)             & 15                & 3.0\% \\ \hline
Disable Interfering Factors (DIF) & \textit{Disable Markdown All in One, and it will work.} (Discussion \#10203)                          & 15               & 3.0\% \\ \hline
Restart Runtime Environment (RRE)        & \textit{Restart your machine and IDE as a quick fix.} (Discussion \#17638) & 11                &2.2\% \\ \hline
% MIW (Modify the Input Way)         &\textit{provide it with enough context.} (Discussion \# 7516)  & 7             & 1.4\% \\ \hline
Modify the Input Way (MIW)         & \textit{Go on new line after comment, and start typing what you want to create.} (SO \# 70995718)  & 8             & 1.6\% \\ \hline
Install/Update Framework (IUF)  &\textit{Have to install node.js for it to work.} (Discussion \#40300)  & 5    & 1.0\% \\ \hline
% Others             & \textit{You can use shortcut Alt + \ to manually request completions.} (Discussion \#7254)
%                                       & 21                & 4.2\% \\ \hline
Others             & \textit{Installing Win-CA helps hides it again and all looks back to normal.} (SO \#71367058)
                                      & 21                & 4.2\% \\ \hline
\end{tabular}
\end{table*}

\subsubsection{Solutions to Problems Mapping}
% The mapping between problem \textcolor{blue}{categories and} types and solution types with the distribution is shown in Table~\ref{tab:issues-solutions}, using abbreviations to represent each type of solution. For example, ``BFC'' represents \textit{Bug Fixed by Copilot}. 
Table~\ref{tab:issues-solutions} illustrates the mapping relationship of Copilot related problems to their solution types, using abbreviations to represent each type of solution. For example, ``BFC'' represents \textit{Bug Fixed by Copilot}. 
% Table \ref{The types of solutions to the issues when using Copilot} presents the full names for all types of solutions. 
The full names for all types of solutions are provided in the note of Table \ref{tab:issues-solutions}.

For \textit{Operation Issue}, 44.5\% of the cases are effectively addressed. Specifically, \textsc{accessing failure} is mainly addressed by BFC; \textsc{authentication failure} is primarily resolved by BFC, MCS, and RC; \textsc{functionality failure} is commonly fixed by BFC, MCS, and USV; and \textsc{startup issue} is mostly solved by BFC, MCS, USV, and RC.
% A large number of effective solutions are identified from \textsc{authentication failure}, \textsc{functionality failure}, and \textsc{startup issue}. Additionally, BFC, MCS, USV, and RC were the most frequently employed solutions for resolving such problems. 
To address \textsc{installation issue} and \textsc{version control issue}, USV is the primarily employed solution.

For \textit{Compatibility Issue}, effective solutions were identified in 39.3\% of the cases. BFC, MCS, and USV are the commonly used solutions for resolving \textsc{editor/ide compatibility issue}, while BFC, MCS, and USV are mostly employed for addressing \textsc{plug-in compatibility issue}. Moreover, resolving \textsc{keyboard compatibility issue} mainly relies on MCS.

For \textit{Feature Request}, 21.7\% of the cases have effective solutions. Specifically, \textsc{function request} is primarily addressed by MCS and FIC. Two cases of \textsc{ui request} were addressed by MCS. In addition, although the solutions for \textsc{professional copilot version} have not been identified, GitHub announced the availability of Copilot Enterprise on February 27, 2024 \citep{copilotenterprise}, potentially addressing the need for a professional version of Copilot.

We identified effective solutions for 24.1\% of \textit{User Experience Issues}. 
% The solutions for all the four types of problems in this category are quite limited, primarily relying on employing BFC and MCS for solutions. 
\textsc{poor authentication experience} is addressed by BFC, USV, and RC, while \textsc{poor functionality experience} is mainly resolved by MCS. Moreover, BFC is the only solution identified for addressing \textsc{poor performance}.

We only identified five solutions for \textit{Suggestion Content Issue}, accounting for 10.2\% of the total cases. BFC and MIW were employed to address \textsc{nonsensical suggestion} and \textsc{low quality suggestion issues}, and no solutions have been identified for resolving \textsc{insecure suggestion}, \textsc{less efficient suggestion}, \textsc{suggestion with bugs}, \textsc{incomprehensible suggestion}, and \textsc{suggestion with invalid syntax}.

Only 6.7\% of \textit{Copyright and Policy Issues} have corresponding solutions. For \textsc{code copyright issue} and \textsc{code telemetry issue}, MCS is the identified solution.

\begin{table*}[h]
\caption{Mapping between problem types (vertical) and solution types (horizontal)}
\label{tab:issues-solutions}
\begin{adjustbox}{width=\textwidth,center}
\begin{tabular}{clccccccccccc}
                                  & & \multicolumn{1}{l}{\textbf{BFC}}     & \multicolumn{1}{l}{\textbf{MCS}} & \multicolumn{1}{l}{\textbf{USV}} & \multicolumn{1}{l}{\textbf{RC}} & \multicolumn{1}{l}{\textbf{FIC}} & \multicolumn{1}{l}{\textbf{FOI}} & 
                                  \multicolumn{1}{l}{\textbf{DIF}} & \multicolumn{1}{l}{\textbf{RRE}} & \multicolumn{1}{l}{\textbf{MIW}} & \multicolumn{1}{l}{\textbf{IUF}} & \multicolumn{1}{l}{\textbf{Others}} \\
\multirow{4}{*}{\textbf{{\normalsize Compatibility Issue}}} &\cellcolor[HTML]{E0F7E0}Editor/IDE Compatibility Issue       & \cellcolor[HTML]{8EAADB}18 & \cellcolor[HTML]{8EAADB}10        & \cellcolor[HTML]{8EAADB}12          & \cellcolor[HTML]{F2F2F2}0          & \cellcolor[HTML]{D9E2F3}1      & 
\cellcolor[HTML]{D9E2F3}1          & \cellcolor[HTML]{D9E2F3}1      & \cellcolor[HTML]{F2F2F2}0          & \cellcolor[HTML]{D9E2F3}1      & \cellcolor[HTML]{F2F2F2}0          & \cellcolor[HTML]{F2F2F2}0       \\
% &\cellcolor[HTML]{E0F7E0}Framework Compatibility Issue        & \cellcolor[HTML]{F2F2F2}0   & \cellcolor[HTML]{F2F2F2}0        & \cellcolor[HTML]{D9E2F3}1          & \cellcolor[HTML]{F2F2F2}0          & \cellcolor[HTML]{F2F2F2}0      & 
% \cellcolor[HTML]{F2F2F2}0          & \cellcolor[HTML]{F2F2F2}0      & \cellcolor[HTML]{F2F2F2}0          & \cellcolor[HTML]{F2F2F2}0      & \cellcolor[HTML]{F2F2F2}0          & \cellcolor[HTML]{F2F2F2}0       \\
&\cellcolor[HTML]{E0F7E0}Keyboard Compatibility Issue           & \cellcolor[HTML]{F2F2F2}0  & \cellcolor[HTML]{D9E2F3}2        & \cellcolor[HTML]{F2F2F2}0         & \cellcolor[HTML]{F2F2F2}0          & \cellcolor[HTML]{F2F2F2}0      & 
\cellcolor[HTML]{F2F2F2}0          & \cellcolor[HTML]{F2F2F2}0      & \cellcolor[HTML]{F2F2F2}0          & \cellcolor[HTML]{F2F2F2}0      & \cellcolor[HTML]{F2F2F2}0          & \cellcolor[HTML]{F2F2F2}0       \\
&\cellcolor[HTML]{E0F7E0}Plug-in Compatibility Issue      & \cellcolor[HTML]{8EAADB}14   & \cellcolor[HTML]{8EAADB}16        & \cellcolor[HTML]{D9E2F3}1         & \cellcolor[HTML]{F2F2F2}0          & \cellcolor[HTML]{F2F2F2}0      & 
\cellcolor[HTML]{F2F2F2}0          & \cellcolor[HTML]{D9E2F3}3      & \cellcolor[HTML]{F2F2F2}0          & \cellcolor[HTML]{F2F2F2}0      & \cellcolor[HTML]{F2F2F2}0          & \cellcolor[HTML]{D9E2F3}4       \\
\multirow{3}{*}{\textbf{{\normalsize Copyright and Policy Issue}}} &\cellcolor[HTML]{FFFFE0}Code Telemetry Issue                    & \cellcolor[HTML]{F2F2F2}0   & \cellcolor[HTML]{D9E2F3}2        & \cellcolor[HTML]{F2F2F2}0         & \cellcolor[HTML]{F2F2F2}0          & \cellcolor[HTML]{F2F2F2}0      & 
\cellcolor[HTML]{F2F2F2}0          & \cellcolor[HTML]{F2F2F2}0      & \cellcolor[HTML]{F2F2F2}0          & \cellcolor[HTML]{F2F2F2}0      & \cellcolor[HTML]{F2F2F2}0          & \cellcolor[HTML]{F2F2F2}0       \\
& \cellcolor[HTML]{FFFFE0}Code Copyright Issue                    & \cellcolor[HTML]{F2F2F2}0   & \cellcolor[HTML]{D9E2F3}1        & \cellcolor[HTML]{F2F2F2}0         & \cellcolor[HTML]{F2F2F2}0          & \cellcolor[HTML]{F2F2F2}0      & 
\cellcolor[HTML]{F2F2F2}0          & \cellcolor[HTML]{F2F2F2}0      & \cellcolor[HTML]{F2F2F2}0          & \cellcolor[HTML]{F2F2F2}0      & \cellcolor[HTML]{F2F2F2}0          & \cellcolor[HTML]{F2F2F2}0       \\
& \cellcolor[HTML]{FFFFE0}Violation of Marketplace Policy  & \cellcolor[HTML]{F2F2F2}0   & \cellcolor[HTML]{F2F2F2}0        & \cellcolor[HTML]{F2F2F2}0         & \cellcolor[HTML]{F2F2F2}0          & \cellcolor[HTML]{F2F2F2}0      & 
\cellcolor[HTML]{F2F2F2}0          & \cellcolor[HTML]{F2F2F2}0      & \cellcolor[HTML]{F2F2F2}0          & \cellcolor[HTML]{F2F2F2}0      & \cellcolor[HTML]{F2F2F2}0          & \cellcolor[HTML]{F2F2F2}0       \\
\multirow{4}{*}{\textbf{{\normalsize Feature Request}}} &\cellcolor[HTML]{E0F7E0}Function Request                 & \cellcolor[HTML]{F2F2F2}0   & \cellcolor[HTML]{8EAADB}9        & \cellcolor[HTML]{8EAADB}3          & \cellcolor[HTML]{F2F2F2}0          & \cellcolor[HTML]{8EAADB}14       & \cellcolor[HTML]{D9E2F3}1          & \cellcolor[HTML]{F2F2F2}0      & \cellcolor[HTML]{F2F2F2}0          & \cellcolor[HTML]{D9E2F3}1      & \cellcolor[HTML]{F2F2F2}0          & \cellcolor[HTML]{F2F2F2}0       \\
& \cellcolor[HTML]{E0F7E0}Integration Request        & \cellcolor[HTML]{F2F2F2}0   & \cellcolor[HTML]{F2F2F2}0        & \cellcolor[HTML]{F2F2F2}0         & \cellcolor[HTML]{F2F2F2}0          & \cellcolor[HTML]{8EAADB}13      & 
\cellcolor[HTML]{F2F2F2}0          & \cellcolor[HTML]{F2F2F2}0      & \cellcolor[HTML]{F2F2F2}0          & \cellcolor[HTML]{F2F2F2}0      & \cellcolor[HTML]{F2F2F2}0          & \cellcolor[HTML]{F2F2F2}0       \\
& \cellcolor[HTML]{E0F7E0}Professional Copilot Version            & \cellcolor[HTML]{F2F2F2}0   & \cellcolor[HTML]{F2F2F2}0        & \cellcolor[HTML]{F2F2F2}0         & \cellcolor[HTML]{F2F2F2}0          & \cellcolor[HTML]{F2F2F2}0      & 
\cellcolor[HTML]{F2F2F2}0          & \cellcolor[HTML]{F2F2F2}0      & \cellcolor[HTML]{F2F2F2}0          & \cellcolor[HTML]{F2F2F2}0      & \cellcolor[HTML]{F2F2F2}0          & \cellcolor[HTML]{F2F2F2}0       \\
& \cellcolor[HTML]{E0F7E0}UI Request                    & \cellcolor[HTML]{F2F2F2}0  & \cellcolor[HTML]{D9E2F3}2        & \cellcolor[HTML]{F2F2F2}0         & \cellcolor[HTML]{F2F2F2}0          & \cellcolor[HTML]{F2F2F2}0      & 
\cellcolor[HTML]{F2F2F2}0          & \cellcolor[HTML]{F2F2F2}0      & \cellcolor[HTML]{F2F2F2}0          & \cellcolor[HTML]{F2F2F2}0      & \cellcolor[HTML]{F2F2F2}0          & \cellcolor[HTML]{F2F2F2}0       \\
\multirow{7}{*}{\textbf{{\normalsize Suggestion Content Issue}}} &\cellcolor[HTML]{FFFFE0}Insecure Suggestion                 & \cellcolor[HTML]{F2F2F2}0   & \cellcolor[HTML]{F2F2F2}0        & \cellcolor[HTML]{F2F2F2}0         & \cellcolor[HTML]{F2F2F2}0          & \cellcolor[HTML]{F2F2F2}0      & 
\cellcolor[HTML]{F2F2F2}0          & \cellcolor[HTML]{F2F2F2}0      & \cellcolor[HTML]{F2F2F2}0          & \cellcolor[HTML]{F2F2F2}0      & \cellcolor[HTML]{F2F2F2}0          & \cellcolor[HTML]{F2F2F2}0       \\
& \cellcolor[HTML]{FFFFE0}Less Efficient Suggestion                 & \cellcolor[HTML]{F2F2F2}0   & \cellcolor[HTML]{F2F2F2}0        & \cellcolor[HTML]{F2F2F2}0         & \cellcolor[HTML]{F2F2F2}0          & \cellcolor[HTML]{F2F2F2}0      & 
\cellcolor[HTML]{F2F2F2}0          & \cellcolor[HTML]{F2F2F2}0      & \cellcolor[HTML]{F2F2F2}0          & \cellcolor[HTML]{F2F2F2}0      & \cellcolor[HTML]{F2F2F2}0          & \cellcolor[HTML]{F2F2F2}0       \\
& \cellcolor[HTML]{FFFFE0}Low Quality Suggestion                & \cellcolor[HTML]{F2F2F2}0   & \cellcolor[HTML]{F2F2F2}0        & \cellcolor[HTML]{F2F2F2}0         & \cellcolor[HTML]{F2F2F2}0          & \cellcolor[HTML]{F2F2F2}0      & 
\cellcolor[HTML]{F2F2F2}0          & \cellcolor[HTML]{F2F2F2}0      & \cellcolor[HTML]{F2F2F2}0          & \cellcolor[HTML]{D9E2F3}1      & \cellcolor[HTML]{F2F2F2}0          & \cellcolor[HTML]{D9E2F3}2       \\
& \cellcolor[HTML]{FFFFE0}Nonsensical Suggestion                & \cellcolor[HTML]{D9E2F3}2   & \cellcolor[HTML]{F2F2F2}0        & \cellcolor[HTML]{F2F2F2}0         & \cellcolor[HTML]{F2F2F2}0          & \cellcolor[HTML]{F2F2F2}0      & 
\cellcolor[HTML]{F2F2F2}0          & \cellcolor[HTML]{F2F2F2}0      & \cellcolor[HTML]{F2F2F2}0          & \cellcolor[HTML]{D9E2F3}1      & \cellcolor[HTML]{F2F2F2}0          & \cellcolor[HTML]{F2F2F2}0       \\
& \cellcolor[HTML]{FFFFE0}Suggestion with Bugs               & \cellcolor[HTML]{F2F2F2}0   & \cellcolor[HTML]{F2F2F2}0        & \cellcolor[HTML]{F2F2F2}0         & \cellcolor[HTML]{F2F2F2}0          & \cellcolor[HTML]{F2F2F2}0      & 
\cellcolor[HTML]{F2F2F2}0          & \cellcolor[HTML]{F2F2F2}0      & \cellcolor[HTML]{F2F2F2}0          & \cellcolor[HTML]{F2F2F2}0      & \cellcolor[HTML]{F2F2F2}0          & \cellcolor[HTML]{F2F2F2}0       \\
& \cellcolor[HTML]{FFFFE0}Incomprehensible Suggestion         & \cellcolor[HTML]{F2F2F2}0   & \cellcolor[HTML]{F2F2F2}0        & \cellcolor[HTML]{F2F2F2}0         & \cellcolor[HTML]{F2F2F2}0          & \cellcolor[HTML]{F2F2F2}0      & 
\cellcolor[HTML]{F2F2F2}0          & \cellcolor[HTML]{F2F2F2}0      & \cellcolor[HTML]{F2F2F2}0          & \cellcolor[HTML]{F2F2F2}0      & \cellcolor[HTML]{F2F2F2}0          & \cellcolor[HTML]{F2F2F2}0       \\
& \cellcolor[HTML]{FFFFE0}Suggestion with Invalid Syntax          & \cellcolor[HTML]{F2F2F2}0   & \cellcolor[HTML]{F2F2F2}0        & \cellcolor[HTML]{F2F2F2}0         & \cellcolor[HTML]{F2F2F2}0          & \cellcolor[HTML]{F2F2F2}0      & 
\cellcolor[HTML]{F2F2F2}0          & \cellcolor[HTML]{F2F2F2}0      & \cellcolor[HTML]{F2F2F2}0          & \cellcolor[HTML]{F2F2F2}0      & \cellcolor[HTML]{F2F2F2}0          & \cellcolor[HTML]{F2F2F2}0       \\
\multirow{6}{*}{\textbf{{\normalsize Operation Issue}}} & \cellcolor[HTML]{E0F7E0}Accessing Failure          & \cellcolor[HTML]{D9E2F3}8   & \cellcolor[HTML]{D9E2F3}4        & \cellcolor[HTML]{D9E2F3}3         & \cellcolor[HTML]{D9E2F3}4          & \cellcolor[HTML]{D9E2F3}2      & 
\cellcolor[HTML]{D9E2F3}1          & \cellcolor[HTML]{F2F2F2}0      & \cellcolor[HTML]{D9E2F3}2          & \cellcolor[HTML]{F2F2F2}0      & \cellcolor[HTML]{F2F2F2}0          & \cellcolor[HTML]{D9E2F3}2       \\
& \cellcolor[HTML]{E0F7E0}Authentication Failure          & \cellcolor[HTML]{4472C4}31   & \cellcolor[HTML]{8EAADB}15        & \cellcolor[HTML]{D9E2F3}7         & \cellcolor[HTML]{4472C4}30          & \cellcolor[HTML]{D9E2F3}2      & 
\cellcolor[HTML]{D9E2F3}1          & \cellcolor[HTML]{D9E2F3}3      & \cellcolor[HTML]{D9E2F3}1          & \cellcolor[HTML]{F2F2F2}0      & \cellcolor[HTML]{D9E2F3}2          & \cellcolor[HTML]{D9E2F3}3      \\
& \cellcolor[HTML]{E0F7E0}Functionality Failure          & \cellcolor[HTML]{4472C4}27   & \cellcolor[HTML]{4472C4}31        & \cellcolor[HTML]{8EAADB}14         & \cellcolor[HTML]{D9E2F3}8          & \cellcolor[HTML]{F2F2F2}0      & 
\cellcolor[HTML]{D9E2F3}8          & \cellcolor[HTML]{D9E2F3}5      & \cellcolor[HTML]{D9E2F3}2          & \cellcolor[HTML]{D9E2F3}3      & \cellcolor[HTML]{F2F2F2}0          & \cellcolor[HTML]{D9E2F3}5       \\
& \cellcolor[HTML]{E0F7E0}Installation Issue                & \cellcolor[HTML]{D9E2F3}2   & \cellcolor[HTML]{D9E2F3}3        & \cellcolor[HTML]{8EAADB}11         & \cellcolor[HTML]{D9E2F3}1          & \cellcolor[HTML]{F2F2F2}0      & 
\cellcolor[HTML]{F2F2F2}0          & \cellcolor[HTML]{F2F2F2}0      & \cellcolor[HTML]{D9E2F3}1          & \cellcolor[HTML]{F2F2F2}0      & \cellcolor[HTML]{F2F2F2}0          & \cellcolor[HTML]{F2F2F2}0       \\
& \cellcolor[HTML]{E0F7E0}Startup Issue          & \cellcolor[HTML]{4472C4}23   & \cellcolor[HTML]{8EAADB}11        & \cellcolor[HTML]{4472C4}24         & \cellcolor[HTML]{8EAADB}15          & \cellcolor[HTML]{F2F2F2}0      & 
\cellcolor[HTML]{D9E2F3}2          & \cellcolor[HTML]{D9E2F3}3      & \cellcolor[HTML]{D9E2F3}4          & \cellcolor[HTML]{F2F2F2}0      & \cellcolor[HTML]{D9E2F3}3          & \cellcolor[HTML]{D9E2F3}5       \\
& \cellcolor[HTML]{E0F7E0}Version Control Issue          & \cellcolor[HTML]{D9E2F3}4   & \cellcolor[HTML]{D9E2F3}1        & \cellcolor[HTML]{D9E2F3}8         & \cellcolor[HTML]{D9E2F3}1          & \cellcolor[HTML]{F2F2F2}0      & 
\cellcolor[HTML]{F2F2F2}0          & \cellcolor[HTML]{F2F2F2}0      & \cellcolor[HTML]{D9E2F3}1          & \cellcolor[HTML]{F2F2F2}0      & \cellcolor[HTML]{F2F2F2}0          & \cellcolor[HTML]{F2F2F2}0       \\
\multirow{4}{*}{\textbf{{\normalsize User Experience Issue}}} &\cellcolor[HTML]{FFFFE0}Poor Authentication Experience          & \cellcolor[HTML]{D9E2F3}2   & \cellcolor[HTML]{F2F2F2}0        & \cellcolor[HTML]{D9E2F3}1         & \cellcolor[HTML]{D9E2F3}1          & \cellcolor[HTML]{F2F2F2}0      & 
\cellcolor[HTML]{F2F2F2}0          & \cellcolor[HTML]{F2F2F2}0      & \cellcolor[HTML]{F2F2F2}0          & \cellcolor[HTML]{F2F2F2}0      & \cellcolor[HTML]{F2F2F2}0          & \cellcolor[HTML]{F2F2F2}0       \\
& \cellcolor[HTML]{FFFFE0}Poor Functionality Experience          & \cellcolor[HTML]{F2F2F2}0   & \cellcolor[HTML]{D9E2F3}3        & \cellcolor[HTML]{D9E2F3}1         & \cellcolor[HTML]{F2F2F2}0          & \cellcolor[HTML]{F2F2F2}0      & 
\cellcolor[HTML]{F2F2F2}0          & \cellcolor[HTML]{F2F2F2}0      & \cellcolor[HTML]{F2F2F2}0          & \cellcolor[HTML]{D9E2F3}1      & \cellcolor[HTML]{F2F2F2}0          & \cellcolor[HTML]{F2F2F2}0       \\
& \cellcolor[HTML]{FFFFE0}Poor Performance                  & \cellcolor[HTML]{D9E2F3}3   & \cellcolor[HTML]{F2F2F2}0        & \cellcolor[HTML]{F2F2F2}0         & \cellcolor[HTML]{F2F2F2}0          & \cellcolor[HTML]{F2F2F2}0      & 
\cellcolor[HTML]{F2F2F2}0          & \cellcolor[HTML]{F2F2F2}0      & \cellcolor[HTML]{F2F2F2}0          & \cellcolor[HTML]{F2F2F2}0      & \cellcolor[HTML]{F2F2F2}0          & \cellcolor[HTML]{F2F2F2}0       \\
& \cellcolor[HTML]{FFFFE0}Poor Subscription Experience           & \cellcolor[HTML]{D9E2F3}1   & \cellcolor[HTML]{F2F2F2}0        & \cellcolor[HTML]{F2F2F2}0         & \cellcolor[HTML]{F2F2F2}0          & \cellcolor[HTML]{F2F2F2}0      & 
\cellcolor[HTML]{D9E2F3}1          & \cellcolor[HTML]{F2F2F2}0      & \cellcolor[HTML]{F2F2F2}0          & \cellcolor[HTML]{F2F2F2}0      & \cellcolor[HTML]{F2F2F2}0          & \cellcolor[HTML]{F2F2F2}0       \\
\end{tabular}
\end{adjustbox}
\begin{minipage}{18cm} 
\vspace{0.1cm}
\vspace{0.1cm}
\scriptsize  \textbf{Full names of each solution type:} BFC: \textit{Bug Fixed by Copilot}; MCS: \textit{Modify Configuration/Setting}; USV: \textit{Use Suitable Version}; RC: \textit{Reinstall/Restart/Reauthorize Copilot}; FIC: \textit{Feature Implemented by Copilot}; FOI: \textit{Follow Official Instruction}; DIF: \textit{Disable Interfering Factors}; RRE: \textit{Restart Runtime Environment}; MIW: \textit{Modify the Input Way}; IUF: \textit{Install/Update Framework}.
\end{minipage}
\end{table*}

\subsubsection{\textbf{Interpretation}}
\textbf{Frequency of Solutions}: BFC is the most commonly employed solution for addressing problems related to Copilot usage, which is reasonable since \textit{Copilot Internal Error} is identified as the most common cause of Copilot usage problems, indicating that many problems of Copilot cannot be resolved by users. MCS, USV, and RC are frequently employed solutions when users attempt to resolve Copilot related problems by themselves. Typically, users can obtain relevant experience and knowledge in resolving Copilot usage problems by these three methods on public Q\&A platforms (e.g., GitHub Issues, GitHub Discussions, SO). FIC ranks as the fifth most frequently used solution, reflecting the expansion and improvement of Copilot features to match the requirements of a large user community. The remaining six types of solution account for only 11.8\% of the total number of solutions, but can still provide valuable experiences for users who face specific Copilot related problems. For instance, DIF can effectively resolve the conflicts between the plug-in ``Markdown All in One'' and ``Copilot'', saving lots of time for the users who experience the same problem.

\textbf{Mapping of Solutions to Problems}: For \textit{Operation Issue} and \textit{Compatibility Issue}, a great number of effective solutions have been identified. This is partially due to the high number of \textit{Operation Issue} and \textit{Compatibility Issue}, and their direct impact on the proper functioning of Copilot. As a result, both the Copilot team and users are inclined to promptly address these two categories of Copilot related problems. Besides the errors in Copilot server that can be addressed by BFC, some \textit{Operation Issues} and \textit{Compatibility Issues} are attributed to the Copilot running environments of users. Therefore, users can address the problems in these two categories through MCS, USV, and RC.
The relatively limited number of solutions for \textit{Suggestion Content Issue} reflects the lack of effective methods for users to adjust the code suggested by Copilot. \textit{Feature Request} and \textit{User Experience Issue} usually require a new release of Copilot by the Copilot team to expand and optimize the features of Copilot, making it difficult to meet the expectations of some users in a short term. Additionally, we found that MCS is frequently utilized by users when trying to get the functionality they want without waiting the new release of Copilot. Fewer solutions were identified for \textit{Copyright and Policy Issue}, indicating that it is important to provide satisfactory solutions to address the concerns of code leakage.

\section{Implications}
\label{sec:implication}
\subsection{Implications for the Copilot Users}
\begin{tcolorbox}[arc=0mm,width=\columnwidth,
                  top=1mm,left=1mm,  right=1mm, bottom=1mm,
                  boxrule=.2pt]
\textbf{Implication 1}. Seeking inspiration from Copilot's code suggestions rather than relying on them can lead to better control over your coding tasks.
\end{tcolorbox}
%\textbf{Seeking inspiration from Copilot's code suggestions rather than relying on them can lead to better control over your coding tasks.} 
In \textit{Feature Request}, we have observed that users frequently present their wish to accept code suggestions word-by-word or line-by-line when reviewing them. As a user complained in an SO post, ``\textit{Copilot inputs tons of code when people just want a tiny part}'' (SO \#74091857). Users' desire for such requests may reflect their concerns about losing control over their coding tasks because Copilot sometimes is too ``productive'' for them. In the experiment conducted by \cite{vaithilingam2022Expectation}, five participants perceived the same worry when they failed to understand the code suggested by Copilot. Thus, finding a right balance between leveraging Copilot's productivity and maintaining developers' command over their code is important for Copilot users. As pointed out by \cite{ziegler2024measuring}, the value of Copilot lies in whether the suggestions serve as a useful starting point for further development. Therefore, a useful position of Copilot would be a tool that provides coding insights, rather than a complete replacement for developers in coding tasks. Before using Copilot for coding, users should ideally have a clear understanding of their code structure. Then, by guiding Copilot to provide code suggestions, they can further optimize the code implementation. This way ensures that users maintain control over their code while incorporating the inspiring suggestions provided by Copilot, keeping the role of users as a real ``pilot'' of the coding tasks.

\begin{tcolorbox}[arc=0mm,width=\columnwidth,
                  top=1mm,left=1mm,  right=1mm, bottom=1mm,
                  boxrule=.2pt]
\textbf{Implication 2}. Review the code suggestions provided by Copilot before accepting them to prevent introducing quality issues into your project.
\end{tcolorbox}
%\textbf{Review the code suggestions provided by Copilot before accepting them to prevent introducing quality issues into your project.} 
According to the results of RQ1, the emergence of \textsc{insecure suggestion}, \textsc{less efficient suggestion}, and \textsc{suggestion with invalid syntax} indicates potential quality issues in the code suggested by Copilot. The code quality of Copilot generated code has been a critical topic for the Copilot team and many researchers (e.g.,~\cite{yetistiren2022assessing}, \cite{siddiq2022empirical}). Since the outputs of LLMs are unpredictable and the code snippets used to train Copilot potentially contain quality issues, it is unrealistic to expect Copilot to consistently produce quality-assured code under all circumstances. In our study, we identified only a few issues related to the quality of Copilot-generated code. One possible reason is that developers may not be attentive enough and may lack the experience to identify quality issues such as code security in Copilot generated code. According to research conducted by \cite{fu2023security}, 29.8\% of the 452 code snippets produced by Copilot and collected from GitHub contain security issues spanning 38 Common Weakness Enumeration (CWE) \citep{cwe2023} categories. These code snippets were produced in practical software development rather than controlled environments, which reveals the severe security challenges of the code produced by Copilot in real-world scenarios. Users need to have the necessary security awareness and skills when dealing with the code suggestions generated by Copilot. Additionally, other quality issues such as code smells \citep{zhang2024copilot} and code efficiency \citep{nguyen2022evalution} present in Copilot code suggestions also require users to exercise critical evaluation and judgment. We noticed that users may find it challenging to fully understand the code suggestions provided by Copilot in \textsc{incomprehensible suggestion}, leading to difficulties in recognizing the quality issues introduced by Copilot. Therefore, it is essential to carefully examine whether there are any quality issues with the code suggestions from Copilot before accepting them. If users want to utilize code suggestions from Copilot in critical software projects, rigorous code reviews and thorough testing are necessary.

\begin{table*}[h]
\caption{Mapping between problem categories (vertical) and IDEs/editors (horizontal)}
\label{tab:issues-IDE}
\begin{adjustbox}{width=\textwidth,center}
\begin{tabular}{lcccccccccc}
                                  & \multicolumn{1}{l}{\textbf{VSCode}}     & \multicolumn{1}{l}{\textbf{Neovim}} & \multicolumn{1}{l}{\textbf{Visual Studio}} & \multicolumn{1}{l}{\textbf{IDEA}} & \multicolumn{1}{l}{\textbf{PyCharm}} & \multicolumn{1}{l}{\textbf{VSCodium}} & 
                                  \multicolumn{1}{l}{\textbf{Emacs}} & \multicolumn{1}{l}{\textbf{PhpStorm}} & \multicolumn{1}{l}{\textbf{Vim}} & \multicolumn{1}{l}{\textbf{Others}} \\
\textbf{{\normalsize Compatibility Issue}} & \cellcolor[HTML]{4472C4}130        & \cellcolor[HTML]{8EAADB}24          & \cellcolor[HTML]{BCCDE8}16          & \cellcolor[HTML]{D9E2F3}5      & 
\cellcolor[HTML]{D9E2F3}3          & \cellcolor[HTML]{D9E2F3}2      & \cellcolor[HTML]{D9E2F3}1          & \cellcolor[HTML]{D9E2F3}2      & \cellcolor[HTML]{D9E2F3}2          & \cellcolor[HTML]{D9E2F3}10       \\
\textbf{{\normalsize Copyright and Policy Issue}} & \cellcolor[HTML]{D9E2F3}6        & \cellcolor[HTML]{F2F2F2}0          & \cellcolor[HTML]{D9E2F3}1          & \cellcolor[HTML]{F2F2F2}0      & 
\cellcolor[HTML]{F2F2F2}0          & \cellcolor[HTML]{F2F2F2}0      & \cellcolor[HTML]{D9E2F3}1          & \cellcolor[HTML]{F2F2F2}0      & \cellcolor[HTML]{F2F2F2}0          & \cellcolor[HTML]{D9E2F3}1       \\
\textbf{{\normalsize Feature Request}} & \cellcolor[HTML]{8EAADB}63        & \cellcolor[HTML]{BCCDE8}20          & \cellcolor[HTML]{BCCDE8}14          & \cellcolor[HTML]{D9E2F3}7      & 
\cellcolor[HTML]{D9E2F3}3          & \cellcolor[HTML]{F2F2F2}0      & \cellcolor[HTML]{D9E2F3}5          & \cellcolor[HTML]{D9E2F3}2      & \cellcolor[HTML]{D9E2F3}5          & \cellcolor[HTML]{BCCDE8}14       \\
\textbf{{\normalsize Suggestion Content Issue}} & \cellcolor[HTML]{D9E2F3}9        & \cellcolor[HTML]{D9E2F3}2          & \cellcolor[HTML]{D9E2F3}5          & \cellcolor[HTML]{F2F2F2}0      & 
\cellcolor[HTML]{D9E2F3}1          & \cellcolor[HTML]{F2F2F2}0      & \cellcolor[HTML]{F2F2F2}0          & \cellcolor[HTML]{F2F2F2}0      & \cellcolor[HTML]{F2F2F2}0          & \cellcolor[HTML]{F2F2F2}0       \\
\textbf{{\normalsize Operation Issue}} & \cellcolor[HTML]{4472C4}365        & \cellcolor[HTML]{8EAADB}63          & \cellcolor[HTML]{8EAADB}59          & \cellcolor[HTML]{8EAADB}40      & 
\cellcolor[HTML]{8EAADB}36          & \cellcolor[HTML]{BCCDE8}14      & \cellcolor[HTML]{D9E2F3}6          & \cellcolor[HTML]{BCCDE8}11      & \cellcolor[HTML]{D9E2F3}10          & \cellcolor[HTML]{8EAADB}41       \\
\textbf{{\normalsize User Experience Issue}} & \cellcolor[HTML]{8EAADB}27        & \cellcolor[HTML]{D9E2F3}1          & \cellcolor[HTML]{F2F2F2}0          & \cellcolor[HTML]{D9E2F3}4      & 
\cellcolor[HTML]{D9E2F3}2          & \cellcolor[HTML]{F2F2F2}0      & \cellcolor[HTML]{D9E2F3}1          & \cellcolor[HTML]{D9E2F3}3      & \cellcolor[HTML]{F2F2F2}0          & \cellcolor[HTML]{D9E2F3}2       \\
\end{tabular}
\end{adjustbox}
\end{table*}

\begin{tcolorbox}[arc=0mm,width=\columnwidth,
                  top=1mm,left=1mm,  right=1mm, bottom=1mm,
                  boxrule=.2pt]
\textbf{Implication 3}. It is advisable to utilize IDEs or code editors that are officially supported by Copilot.
\end{tcolorbox}
%\textbf{It is advisable to utilize IDEs or code editors that are officially supported by Copilot.} 
Table \ref{tab:issues-IDE} presents the correlation between the number of Copilot usage problems and the IDEs or code editors that users predominantly utilize. 
We found that discussions regarding Copilot usage problems mainly occurred on mainstream coding development platforms such as VSCode, Neovim, Visual Studio, IDEA, PyCharm. However, the results in Table \ref{tab:issues-IDE} do not imply that using a particular IDE or code editor will result in encountering a large number of specific category of Copilot-related problems. Higher numbers often indicate that such problems have been widely discussed and even already addressed. For instance, users of VSCode reported many \textit{Operational Issues} that have been resolved through discussions. This does not necessarily mean that using Copilot on VSCode will lead to encountering a significant number of \textit{Operation Issues}. Besides, the Copilot team also actively provides solutions to the queries arising on the platforms officially supported by Copilot. As a result, users utilizing Copilot on these IDEs or code editors are more likely to resolve usage problems by finding existing solutions or reporting them on public Q\&A platforms. On the contrary, we found that \textit{Unsupported Platform} ranks as the fourth most frequent cause leading to Copilot usage problems. Based on the results of RQ2 (see Table \ref{tab:issues-causes}), it is also one of the primary reasons causing various types of \textit{Operation Issue}, which suggests that users may encounter unforeseen problems when attempting to use Copilot on the platforms that are not supported yet. These problems are often challenging to resolve. In our study, none of the 31 issues associated with \textit{Unsupported Platform} has been resolved effectively. Therefore, users should prioritize using Copilot on officially supported platforms to ensure an optimal experience.

\subsection{Implications for the Copilot Team}
\begin{tcolorbox}[arc=0mm,width=\columnwidth,
                  top=1mm,left=1mm,  right=1mm, bottom=1mm,
                  boxrule=.2pt]
\textbf{Implication 4}. Provide more customization options to allow users to tailor the behavior of Copilot to align with their own workflow.
\end{tcolorbox}
%\textbf{Provide more customization options to allow users to tailor the behavior of Copilot to align with their own workflow.} 
Among the 114 \textsc{function requests}, we identified 52 instances of such requests to customize the behavior of Copilot in various aspects, accounting for approximately 50\%. Some common requests are \textit{specifying the file types or workspace in which Copilot automatically runs} (11), \textit{modifying the shortcut keys for accepting suggestions} (10), \textit{accepting code suggestions line-by-line or word-by-word} (9), \textit{preventing Copilot from generating certain types of suggestions (e.g., file paths, comments)} (3), and \textit{configuring text color and fonts} (3). Our previous study on the expected feature of Copilot also indicates that the functionalities provided by Copilot have not yet fully met the requirements of users for flexible utilization of the tool \citep{zhang2023demystifying}. Additionally, according to some of the identified \textsc{poor functionality experience} (e.g., perceiving the auto-suggestions of Copilot as distracting, which is also mentioned in the study of \cite{bird2022taking}), we can discern the demand for customizing the behavior of Copilot. It is believed that the extent to which the behavior of Copilot can adapt well to user coding habits is a vital factor in their decision to use Copilot. Therefore, providing flexible and user-friendly customization options is highly beneficial.

\begin{tcolorbox}[arc=0mm,width=\columnwidth,
                  top=1mm,left=1mm,  right=1mm, bottom=1mm,
                  boxrule=.2pt]
\textbf{Implication 5}. Provide more ways to control the content generated by Copilot.
\end{tcolorbox}
%\textbf{Provide more ways to control the content generated by Copilot.} 
According to Table \ref{tab:issues-solutions}, 
%it can be observed that the majority of solutions are aimed at addressing \textit{Operation Issue} and \textit{Compatibility Issue}, while 
there is only a small number of solutions for \textit{Suggestion Content Issue}. Out of the 59 \textit{Suggestion Content Issues}, only five solutions were identified, indicating that users may find it challenging to provide ideal solutions for the problems of content suggested by Copilot. The output of Copilot is inherently unpredictable, and users have limited ways to control the code generated by Copilot besides modifying code context or code comments per se. 
% Therefore additional methods are required for addressing \textit{Suggestion Content Issue}, for instance, allowing developers to interact with Copilot and iterate the generated code till the code reaches the expectation of developers.
Based on our observation, features such as \textit{allowing users to define code styles and conventions}, and \textit{choosing multiple files as the context for Copilot to generate code suggestions} are worth trying.

% \textbf{Enhance compatibility across various IDEs and editors and simplify the configuration of Copilot.} According to the results of RQ1 and RQ2, \textit{Compatibility Issue} is the second-largest category, and \textit{Editor/IDE Compatibility Issue} is one of the main causes that leads to \textit{Operation Issue}. From the perspective of users, we also have observed lots of discussions related to configuration and settings of Copilot, which makes \textit{Modify Configuration/Setting} the second most frequently employed solution. Additionally, \textit{Improper Configuration/Setting} is the fifth most common cause of problems related to Copilot usage. Based on these findings, we believe that enhancing compatibility and simplifying the configuration process of Copilot for users can significantly improve their experience. Therefore, the Copilot team may offer more detailed installation and configuration guidelines, provide user-friendly configuration options.
% % , and perform regular update and maintenance.

\begin{tcolorbox}[arc=0mm,width=\columnwidth,
                  top=1mm,left=1mm,  right=1mm, bottom=1mm,
                  boxrule=.2pt]
\textbf{Implication 6}. Simplify the configuration of Copilot and provide support for more IDEs and code editors.
\end{tcolorbox}
%\textbf{Simplify the configuration of Copilot and provide support for more IDEs and code editors.} 
According to the results of RQ1 and RQ2, \textit{Compatibility Issue} is the second-largest category, and \textit{Editor/IDE Compatibility Issue} is one of the main causes that leads to many \textit{Operation Issues}. From the perspective of users, we also have observed lots of discussions related to configuration and settings of Copilot, which makes \textit{Modify Configuration/Setting} the second most frequently employed solution. Additionally, \textit{Improper Configuration/Setting} is the fifth most common cause of problems related to Copilot usage. Based on these findings, we believe that simplifying the configuration process of Copilot for users can significantly improve their experience. For example, the Copilot team may offer more detailed installation and configuration guidelines, provide user-friendly configuration options. Moreover, we have identified 71 \textsc{integration requests}, indicating a significant number of users expressing a desire for Copilot compatibility across a broader range of IDEs and code editors.
% , and perform regular update and maintenance.

% \textbf{Improve the quality of Copilot generated code.} In \textit{Suggestion Content Issues}, the predominant types are \textsc{low quality suggestion} (27) and \textsc{nonsensical suggestion} (13). The experiment by Imai \textit{et al.} found that, in comparison to human pair programming, Copilot, while is capable of generating a significant amount of code, also led to more code deletions during testing, highlighting the need for improvement in Copilot's code quality \citep{imai2022github}. Bird \textit{et al.} observed that Copilot occasionally offers peculiar and nonsensical code suggestions, as reported by users, some of which may include personal information \citep{bird2022taking}. Furthermore, although \textsc{insecure suggestion} and \textsc{less effective suggestion} each have only two instances, we believe this is primarily due to users encountering difficulty in detecting issues of these kinds and being less inclined to report them.
% % Pearce \textit{et al.} \cite{hammond2022asleep} found that out of the 1,689 code snippets generated by Copilot, 40\% were vulnerable. 
% In the experiment of Vaithilingam \textit{et al.}, some participants mentioned that they prefer involving Copilot only in simple programming tasks due to the concerns about introducing subtle code quality issues or code smells~\citep{vaithilingam2022Expectation}.
% Given the successive iterations of Copilot, it becomes imperative to conduct regular assessments of the quality of Copilot's generated code.

\begin{tcolorbox}[arc=0mm,width=\columnwidth,
                  top=1mm,left=1mm,  right=1mm, bottom=1mm,
                  boxrule=.2pt]
\textbf{Implication 7}. Increase the diversity of the code suggested by Copilot while improving their quality.
\end{tcolorbox}
%\textbf{Increase the diversity of the code suggested by Copilot while improving their quality.} 
In \textit{Suggestion Content Issues}, the predominant types are \textsc{low quality suggestion} and \textsc{nonsensical suggestion}. \cite{bird2022taking} also observed that Copilot occasionally offers peculiar and nonsensical code suggestions as reported by the Copilot users.  
% The experiment by Imai \textit{et al.} found that, in comparison to human pair programming, Copilot, while is capable of generating a significant amount of code, also led to more code deletions during testing, highlighting the need for improvement in Copilot's code quality \citep{imai2022github}. Furthermore, although \textsc{insecure suggestion} and \textsc{less effective suggestion} each have only two instances, we believe this is primarily due to users encountering difficulty in detecting issues of these kinds and being less inclined to report them.
% Pearce \textit{et al.} \cite{hammond2022asleep} found that out of the 1,689 code snippets generated by Copilot, 40\% were vulnerable. 
% In the experiment of Vaithilingam \textit{et al.}, some participants mentioned that they prefer involving Copilot only in simple programming tasks due to the concerns about introducing subtle code quality issues or code smells~\citep{vaithilingam2022Expectation}.
Other prior studies (e.g., \cite{vaithilingam2022Expectation}, \cite{hammond2022asleep}) also indicated the problems with the code suggestions provided by Copilot in terms of code quality. However, compared to the issues of specific code content suggested by Copilot, we found that users are more inclined to discuss situations where Copilot fails to provide valuable code suggestions as problems on public Q\&A platforms. This tendency actually reflects one of the primary purposes for which many users utilize Copilot: they hope that Copilot will inspire their ideas in coding. Observations from some instances of \textit{User Experience Issue} have shown how users can be disappointed when all 10 code suggestions provided by Copilot lack diversity. In fact, in addition to the improvement in code quality, diverse coding suggestions are essential for satisfying the needs of users for more inspirational code.

\begin{table*}[h]
\caption{Mapping between problem categories (vertical) and programming languages (horizontal)}
\label{tab:issues-languages}
\begin{adjustbox}{width=0.8\textwidth,center}
\begin{tabular}{lccccccccc}
                                  & \multicolumn{1}{l}{\textbf{Python}} & \multicolumn{1}{l}{\textbf{C\#}} & \multicolumn{1}{l}{\textbf{JavaScript}} & \multicolumn{1}{l}{\textbf{Php}} & \multicolumn{1}{l}{\textbf{Java}} & 
                                  \multicolumn{1}{l}{\textbf{TypeScript}} & \multicolumn{1}{l}{\textbf{C++}} & \multicolumn{1}{l}{\textbf{Go}} & \multicolumn{1}{l}{\textbf{Others}} \\
\textbf{{\normalsize Compatibility Issue}} & \cellcolor[HTML]{8EAADB}13        & \cellcolor[HTML]{D9E2F3}8          & \cellcolor[HTML]{D9E2F3}8          & \cellcolor[HTML]{D9E2F3}2      & 
\cellcolor[HTML]{D9E2F3}1          & \cellcolor[HTML]{D9E2F3}5      & \cellcolor[HTML]{D9E2F3}3          & \cellcolor[HTML]{D9E2F3}1      & \cellcolor[HTML]{D9E2F3}1       \\
\textbf{{\normalsize Copyright and Policy Issue}} & \cellcolor[HTML]{F2F2F2}0        & \cellcolor[HTML]{D9E2F3}1          & \cellcolor[HTML]{D9E2F3}1          & \cellcolor[HTML]{F2F2F2}0      & 
\cellcolor[HTML]{F2F2F2}0          & \cellcolor[HTML]{F2F2F2}0      & \cellcolor[HTML]{F2F2F2}0          & \cellcolor[HTML]{F2F2F2}0      & \cellcolor[HTML]{F2F2F2}0        \\
\textbf{{\normalsize Feature Request}} & \cellcolor[HTML]{D9E2F3}8        & \cellcolor[HTML]{D9E2F3}8          & \cellcolor[HTML]{D9E2F3}3          & \cellcolor[HTML]{D9E2F3}2      & 
\cellcolor[HTML]{D9E2F3}1          & \cellcolor[HTML]{F2F2F2}0      & \cellcolor[HTML]{D9E2F3}2          & \cellcolor[HTML]{D9E2F3}1      & \cellcolor[HTML]{D9E2F3}4          \\
\textbf{{\normalsize Suggestion Content Issue}} & \cellcolor[HTML]{D9E2F3}10        & \cellcolor[HTML]{D9E2F3}9          & \cellcolor[HTML]{D9E2F3}5          & \cellcolor[HTML]{D9E2F3}1      & 
\cellcolor[HTML]{D9E2F3}1          & \cellcolor[HTML]{D9E2F3}4      & \cellcolor[HTML]{D9E2F3}2          & \cellcolor[HTML]{D9E2F3}1      & \cellcolor[HTML]{D9E2F3}4          \\
\textbf{{\normalsize Operation Issue}} & \cellcolor[HTML]{4472C4}49        & \cellcolor[HTML]{8EAADB}13          & \cellcolor[HTML]{8EAADB}14          & \cellcolor[HTML]{8EAADB}12      & 
\cellcolor[HTML]{8EAADB}15          & \cellcolor[HTML]{D9E2F3}10      & \cellcolor[HTML]{D9E2F3}5          & \cellcolor[HTML]{D9E2F3}4      & \cellcolor[HTML]{D9E2F3}7          \\
\textbf{{\normalsize User Experience Issue}} & \cellcolor[HTML]{D9E2F3}5        & \cellcolor[HTML]{8EAADB}0          & \cellcolor[HTML]{D9E2F3}1          & \cellcolor[HTML]{D9E2F3}3      & 
\cellcolor[HTML]{F2F2F2}0          & \cellcolor[HTML]{D9E2F3}1      & \cellcolor[HTML]{D9E2F3}1          & \cellcolor[HTML]{F2F2F2}0      & \cellcolor[HTML]{F2F2F2}0          \\
\end{tabular}
\end{adjustbox}
\end{table*}
\begin{tcolorbox}[arc=0mm,width=\columnwidth,
                  top=1mm,left=1mm,  right=1mm, bottom=1mm,
                  boxrule=.2pt]
\textbf{Implication 8}. When using popular programming languages, user experience with Copilot shows no noticeable differences. However, additional enhancement may be necessary for more niche or domain-specific languages.
\end{tcolorbox}
Table \ref{tab:issues-languages} presents the correlation between the number of Copilot usage problems and the programming languages that users predominantly utilize. Python, C\#, JavaScript, and Java are the most frequently used programming languages. The official claim from the Copilot team is that Copilot can support almost all programming languages \citep{githubcopilot}. Based on our observations of these reported problems, no commonly used programming language consistently exhibited specific categories of problems for users. Users of these mainstream languages have not indicated significant differences in user experience and the quality of code suggestions when using different languages. However, We found that some users expressed a need for enhancements or support for niche and domain-specific languages when using Copilot. For instance, a user expressed disappointment with the code suggested by Copilot for the Nim language (Discussion \#7180), while another user requested support for Classic ASP and VBScript (Discussion \#53154). Due to varying training data across different programming languages, users may have a suboptimal experience when using less common languages. Therefore, we suggest that the Copilot team collect more relevant training data and further train Copilot for these programming languages to improve the user experience.

\begin{tcolorbox}[arc=0mm,width=\columnwidth,
                  top=1mm,left=1mm,  right=1mm, bottom=1mm,
                  boxrule=.2pt]
\textbf{Implication 9}. Consider intellectual property and copyright when gathering training data for Copilot and giving code suggestions.
\end{tcolorbox}
%\textbf{Consider intellectual property and copyright when gathering training data for Copilot and giving code suggestions.} 
Since the announcement of Copilot, there has been ongoing controversy regarding its implications for intellectual property and copyright. In our dataset, the number of \textit{Copyright and Policy Issue} is higher than we expected, and we observed many related concerns from both users and repository owners during the data extraction process. For the owners of the repositories hosted on GitHub, the primary concern is whether their code has been or will be used illegally to train Copilot. Copilot users are concerned about whether the sensitive code context provided for generating code suggestions would be collected and stored by Copilot. In \textit{Feature Request}, we noticed that some users call for Copilot to introduce new features that prevent \textsc{code telemetry issue}. In addition, \cite{bird2022taking} found some discussions between Copilot users about how copyright applied to code suggestions generated by Copilot. According to their findings, Copilot users hold different views about how to license the code generated by Copilot and whether the code generated by Copilot should be capable of copyright protection.
% The goal of our research is not to provide an evaluation of such problems or the non-open source nature of Copilot. 
Issues related to intellectual property and copyright when using Copilot should be seriously taken into account by the Copilot team to prevent potential impacts on the development of Copilot and user experience. As \cite{akbar2023ethical} pointed out, intellectual property, legal compliance, and privacy are critical ethical principles when developing and using LLM-driven tools like Copilot. Therefore, we contend that the Copilot team should take measures to address these problems, providing high-quality code generation services while protecting user privacy and intellectual property. Besides, we noticed that the Copilot team has published ``GitHub Copilot Trust Center'' \citep{copilottrustcenter} to help answer questions regarding privacy, security, and intellectual property. We plan to further explore the role of this center in addressing user concerns regarding the \textit{Copyright and Policy Issue} in our future work.

\subsection{Implications for Researchers}
\begin{tcolorbox}[arc=0mm,width=\columnwidth,
                  top=1mm,left=1mm,  right=1mm, bottom=1mm,
                  boxrule=.2pt]
\textbf{Implication 10}. The use of Copilot may alter the coding process and increase the time cost of verifying code suggestions, making code explanation highly important.
\end{tcolorbox}
%\textbf{The use of Copilot may alter the coding process and increase the time cost of verifying code suggestions, making code explanation highly important.} 
In our research, \textsc{incomprehensible suggestion} ranks as the fourth most common \textit{Suggestion Content Issue}, and the code explanation feature is also frequently requested. Some users mentioned that the code suggestions generated by Copilot are excessively long, resulting in reduced readability. This indicates that when Copilot provides relatively complex suggestions, or when users lack coding experience in a particular domain, understanding the code logic and verifying its correctness can be time-consuming. In that case, users may refrain from adopting code suggestions of Copilot as it makes them feel a loss of control over the coding task, according to the research of \cite{vaithilingam2022Expectation}. The study by \cite{wang2023investigating} also shows that using AI-generated code can lead to substantial review pressure. Therefore, we believe that AI coding tools like Copilot will change the allocation of time spent on various tasks in software development. As expected by some users, the code explanation feature would be helpful. GitHub Copilot Chat now is generally available for organizations and individuals \citep{copilotchat}, which is powered by GPT-4. The chat feature enables Copilot to interact more flexibly with users, providing the potential for implementing functions such as code explanation, code refactoring, and vulnerability detection. For future research, we plan to explore how the chat feature can influence the programming process and enhance efficiency of coding tasks.

\begin{tcolorbox}[arc=0mm,width=\columnwidth,
                  top=1mm,left=1mm,  right=1mm, bottom=1mm,
                  boxrule=.2pt]
\textbf{Implication 11}. To accurately gauge user satisfaction with Copilot, it is essential to consider programming tasks across various application domains, as well as the purposes for which users employ Copilot.
\end{tcolorbox}
%\textbf{To accurately gauge user satisfaction with Copilot, it is essential to consider programming tasks across various application domains, as well as the purposes for which users employ Copilot.} 
We identified relatively few \textit{Suggestion Content Issues}, and one potential reason is that users are less inclined to report specific code-related issues to public Q\&A platforms, indicating a certain degree of tolerance for AI-generated erroneous suggestions \citep{weisz2021Perfection}. Therefore, our study results may not fully reflect user satisfaction with the Copilot generated code content. Users often propose new \textit{Feature Requests} based on their individual needs and coding habits, coupled with their experience of using Copilot. Some of these requests are closely associated with the domains of the applications that Copilot users develop, such as front-end web development and game development. Consequently, user satisfaction levels with Copilot may vary across different application domains. Thus, assessing the capabilities of Copilot through the lens of single programming tasks, such as addressing algorithmic problems or writing SQL statements, may not represent the actual user satisfaction in other application domains. Furthermore, our previous research found that users have diverse objectives when using Copilot \citep{zhang2023demystifying}, directly influencing their expectations. For instance, users looking to quickly grasp new technologies with the assistance of Copilot are more concerned with whether the code suggestions provide helpful programming guidance for subsequent steps; conversely, those who wish to use Copilot for repetitive coding tasks, such as CRUD operations on a database, are more concerned with whether Copilot can generate correct code to reduce the significant effort they spend on fixing similar issues in code suggestions. Therefore, the intended utilization of Copilot also stands as a critical factor influencing the satisfaction of users with Copilot. A focused study on user satisfaction with Copilot based on these two factors will provide an insightful assessment.

\begin{figure}[htbp]
	\includegraphics[width=\linewidth]{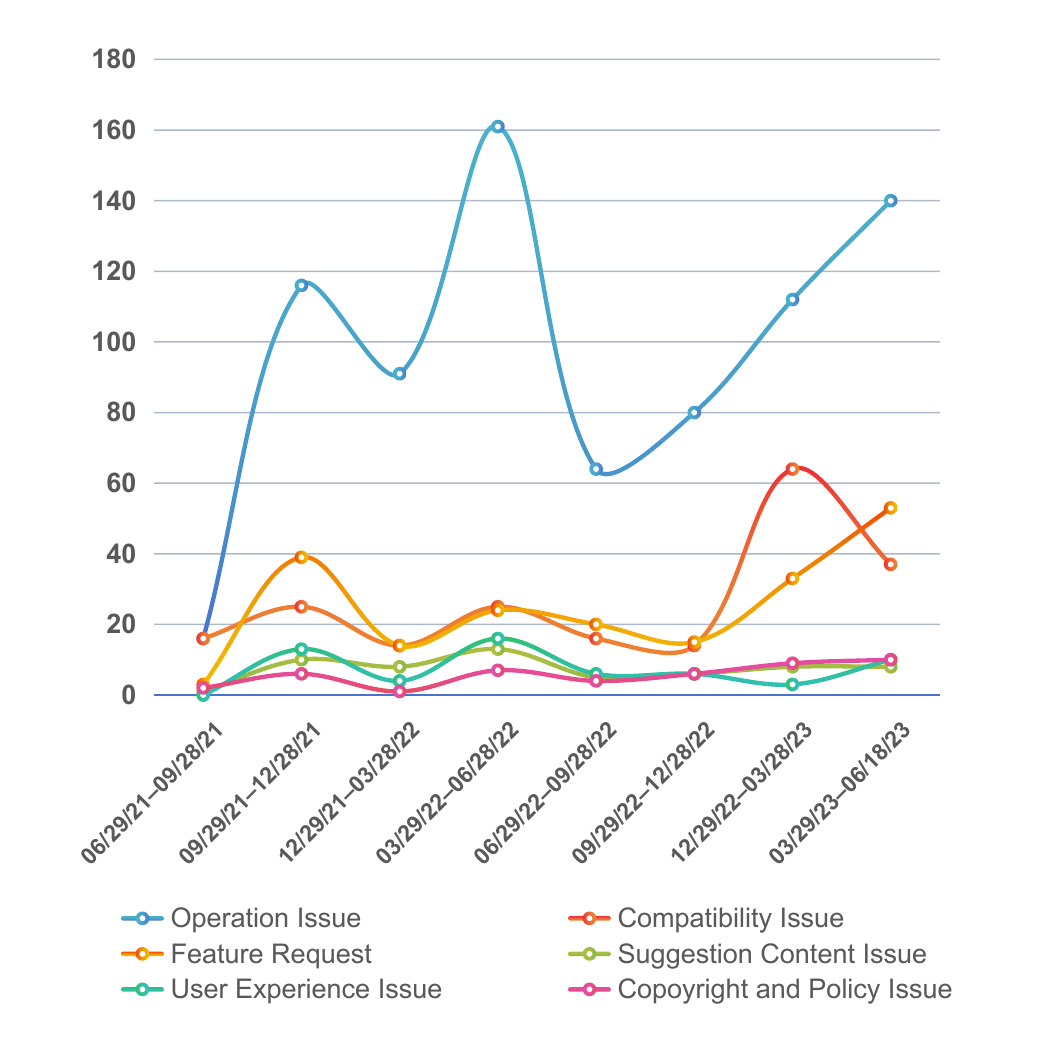}
	\caption{Trends of Copilot Problems}
	\label{fig: Copilot Problems Trends}
\end{figure}

\begin{tcolorbox}[arc=0mm,width=\columnwidth,
                  top=1mm,left=1mm,  right=1mm, bottom=1mm,
                  boxrule=.2pt]
\textbf{Implication 12}. By analyzing the timeline of when problems related to Copilot usage occurred, we found that \textit{Feature Requests} of users increased with the iterations of Copilot.
\end{tcolorbox}

%\textcolor{blue}{\textbf{By taking into account the chronology of when the Copilot usage related problems occurred, we found that the \textit{Feature Requests} of users increased with the iteration of Copilot.} 
Fig. \ref{fig: Copilot Problems Trends} illustrates the variation in the number of six categories of Copilot usage problems every three months. As shown in Fig. \ref{fig: Copilot Problems Trends}, \textit{Operation Issue} outnumbers other categories during almost all the periods, which is reasonable since \textit{Operation Issues} constitute 57.5\% of all Copilot-related problems. Besides, we found that the numbers of \textit{User Experience Issue}, \textit{Suggestion Content Issue}, and \textit{Copy and Policy Issue} have remained relatively stable. However, despite the Copilot team introducing new features and enhancing existing features to meet the requirements of Copilot users, the number of \textit{Feature Requests} has continued to rise since January 2022. Considering the uncertainty of the code generated by Copilot, users are also uncertain about the optimal way to interact with Copilot. Consequently, users provide feedback based on their practical coding experience with Copilot and hope that the optimizations from the Copilot team will enhance their usage experience. Even though the Copilot team continuously introduced new functions to meet user expectations, users can still provide more feedback and come up with additional demands based on their experience with these new features of Copilot. As the version of Copilot evolves, user expectations also evolve correspondingly. Hence, exploring how AI code generation tools such as Copilot can better interact with users could be an interesting and worthy research topic. We argue that researchers should consistently monitor the feedback and trends in \textit{Feature Requests} for Copilot, particularly focusing on the feedback following the introduction of new features, which can guide future enhancements and pinpoint specific areas where Copilot can be better improved.

\section{Threats to Validity}
\label{sec:threats}
The threats to validity are discussed according to the guidelines in \cite{runeson2009guidelines}, and internal validity is not considered, since we did not investigate the relationships between variables and results.

\textbf{Construct validity}: 
%refers to the extent to which a research tool or method can accurately assess the variable or concept being measured, and whether the results obtained are consistent with the RQs, which is a critical aspect of any research study that involves the measurement of variables or concept. 
As the processes of data labelling, data extraction, and data analysis in this study were conducted manually, there is a risk of introducing personal bias. Therefore, we implemented some strategies to enhance the construct validity. 
%Before each formal step of the research process, we conducted pilot experiments to test the validity of the methods and the standards between different researchers. 
In order to reduce this threat, the first and third authors carried out pilot experiments to agree on the criteria for data labelling and data extraction. If any disagreements arose during these processes, the second author was involved to achieve a consensus. The results of data extraction were rechecked by the three authors to ensure accuracy. The data analysis was conducted by the first author. When uncertainties emerged, the first author discussed them with the second and third authors to achieve joint agreement. For the results of the data analysis, the negotiated agreement approach~\citep{campbell2013coding} was employed to address any conflicts.
%The first and third authors would engage in discussions after each step of the pilot experiment and formal research procedures to evaluate the results and ensure consistency with the RQs.

\textbf{External validity}: 
%refers to the generalization of the findings in this study. 
For our research, the primary threat to external validity is the selection of data sources. To maximize external validity,  we chose GitHub Issues, GitHub Discussions, and SO posts as data sources. GitHub Issues is a tool used to report and track software issues, allowing users to report errors, request features, and raise questions to developers. While GitHub Discussions is a new feature on GitHub that aims to provide a more open and organized platform for users to communicate and share insights with other community members. As a popular Q\&A community, Stack Overflow is also a platform for many developers to engage in discussions and share insights regarding Copilot usage. These platforms contain a substantial amount of relevant data, and their data are complementary to each other. Consequently, we were able to collect diverse usage-related data of Copilot from a large number of developers and projects from these three data sources. However, despite all these efforts, we admitted that there may still be relevant data that we missed. We collected the discussions under the ``Copilot'' subcategory of GitHub Community rather than searching for all GitHub discussions by using ``Copilot'' as the keyword. The absence of discussions from other repositories that are potentially related to Copilot could affect the statistics on the number of different categories of Copilot usage problems. Additionally, with the rapid release of the new versions of Copilot, we acknowledge that some new problems related to Copilot usage may emerge, or the problems reported may have been resolved. Hence, there is a need to continuously collect new data in future work to refine our conclusions. Also, industrial surveys can be employed to obtain more information related to the usage of Copilot, such as development contexts and users' coding experience. To validate and replicate our findings and potentially uncover additional insights, we provide our dataset for other researchers \citep{dataset}.
% \textcolor{blue}{We collected the discussions under Copilot category of GitHub Community, rather than searching for all GitHub discussions by using ``Copilot'' as a keyword. Due to the absence of the discussions from other repositories that are potentially related to Copilot, this could affect the statistics on the number of different categories of Copilot usage problems. Additionally, with the rapid version iteration of Copilot, we found that some new problems related to Copilot usage have emerged. For the missing data, we plan to continuously collect it in future work to further refine our conclusions. Also, industrial surveys and interviews will be employed to obtain more information related to the usage of Copilot such as development contexts and coding experience of users.  To validate our findings and potentially uncover additional insights, we  provide our dataset for other researchers \citep{dataset}.} 

\textbf{Reliability}: 
%refers to the degree to which a research method can consistently produce reliable and reproducible results. 
To minimize potential uncertainties arising from the research methodology, we have implemented multiple measures to maximize the reliability of our study. We conducted a pilot labelling to assess the consistency of the two authors prior to the formal data labelling process. The Cohen's Kappa coefficients of the three pilot labelling processes are 0.824, 0.834, and 0.806, indicating good agreement between the authors. Throughout the data labelling, extraction, and analysis process, we thoroughly discussed and resolved any inconsistencies within the team to ensure the consistency and accuracy of the result. Furthermore, we have made available the dataset of the study~\citep{dataset} to enable other researchers to validate our findings.

\section{Related Work}
\label{sec:relatedWork}
\subsection{Evaluating the Quality of Code Generated by Copilot}
Several studies focused on various aspects of code quality generated by Copilot. \cite{siddiq2022empirical} analyzed the prevalence of code smells in the datasets used to train code generation tools like Copilot, and observed the presence of 18 types of code smells in the suggestions provided by Copilot. \cite{yetistiren2022assessing} evaluate the validity, correctness, and efficiency of the Copilot generated code, and their results indicate that Copilot is capable of generating valid code with a success rate of 91.5\%. \cite{hammond2022asleep} prompted Copilot to generate code related to high-risk network security vulnerabilities in order to investigate the conditions that may lead Copilot to suggest insecure code. They found that vulnerabilities were present in 40\% of the cases. \cite{nguyen2022evalution} conducted a study using 33 LeetCode problems to evaluate the correctness and comprehensibility of Copilot in four different programming languages. They found that Copilot's suggestions had low cyclomatic and cognitive complexity and did not show significant differences across programming languages. \cite{dakhel2022asset} investigated the capabilities of Copilot in two different programming tasks and found that it was able to provide solutions for almost all basic algorithmic problems, but some of the solutions were buggy and not replicable. The study conducted by \cite{asare2023githubs} suggests that Copilot won't generate code with the same vulnerabilities as those previously introduced by human developers overall. \cite{sobania2022choose} evaluated the performance of Copilot on standard program synthesis benchmark problems and compared it with results from the genetic programming literature. They found that Copilot demonstrated more mature performance. \cite{mastropaolo2023robustness} aimed to investigate whether different but semantically equivalent natural language descriptions would lead to the generation of the same recommended functions. Their results indicate that differences between semantically equivalent descriptions could affect the correctness of the generated code. \cite{madi2023association} investigated the readability and visual inspection of the code generated by Copilot. They revealed the importance of developers being cautious and vigilant when working with code generation tools such as Copilot. \cite{gustavo2022security} investigated the impact of programming with LLMs that support Copilot. They found that users assisted by LLMs produce critical security bugs at a rate no greater than 10\% more than those not assisted.

\subsection{Copilot's Impact on Practical Development}
Several studies focused on investigating the performance of Copilot in actual software development, as well as the opinions of software practitioners on it. \cite{wang2023practitioners} conducted an interview with 15 practitioners and then surveyed 599 practitioners from 18 IT companies regarding their expectations of code completion. They found that 13\% of the participants had used Copilot as their code completion tool. \cite{jaworski2023study} prepared a survey questionnaire consisting of 18 questions to investigate developers' attitudes toward Copilot. The research findings indicate that most people have a positive attitude towards the tool, but few participants showed concerns about security issues associated with using Copilot. \cite{imai2022github} conducted experiments with 21 participants to compare the effectiveness of Copilot paired with human programmers in terms of productivity and code quality. The results indicate that while Copilot can increase productivity by adding more lines of code, the generated code quality is lower due to the need to remove more lines of code during testing. \cite{shraddha2023grounded} observed 20 participants who collaborated with Copilot to complete different programming tasks in four languages, and found that the interaction with the programming assistant was bimodal in different collaboration mode. \cite{bird2022taking} conducted three studies aimed at understanding how developers utilize Copilot. Their findings suggest that developers spent a lot of time assessing the suggestions generated by Copilot instead of completing their coding tasks. \cite{peng2023impact} presented the results of a controlled experiment using Copilot as an AI collaborative programmer. They found that the experimental group who had access to Copilot completed tasks 55.8\% faster than the control group. \cite{zhang2023demystifying} investigated the programming languages, IDEs, associated technologies, implemented functionalities, advantages, limitations, and challenges when using Copilot. \cite{vaithilingam2022Expectation} conducted a user study involving 24 participants to assess the usability of Copilot and its integration into the programming workflow. They found that while Copilot may not directly enhance the efficiency of completing programming tasks, it serves as a valuable starting point for programmers, saving time spent on searching. \cite{liang2023understanding} conducted a survey among software developers and found that the primary motivation for developers to use AI programming assistants is to reduce keystrokes, complete programming tasks quickly, and recall syntax. However, the impact of using these tools to help generate potential solutions is not significant. \cite{gustavo2022security} analyzed the use of Copilot for programming and compared it with earlier forms of programmer assistance. They also explored potential challenges that could arise when applying LLMs to programming. \cite{ziegler2024measuring} sought to assess the impact of Copilot on user productivity through a case study, aiming to align user perceptions with empirical data. Their research highlights the aspects in which Copilot has enhanced users' coding productivity and how it achieves these improvements.
%\textcolor{blue}{Compared to the existing work, our research collected data from GitHub Issues, Discussions, and SO posts, focusing on the issues users encounter while using Copilot in practical development, as well as their causes and solutions.}
% In contrast to the prior studies, our research distinguishes itself by gathering comprehensive data from GitHub Issues, Discussions, and SO posts. Specifically, our study delves into the problems confronted by users during the practical use of Copilot, exploring both the underlying causes and potential solutions for these problems.

\subsection{Conclusive Summary}
Most of the prior studies utilized controlled experiments or surveys to evaluate the effectiveness of Copilot. Our research is grounded in the perspective of software developers, focusing on the real-world problems they encounter when using Copilot, exploring the underlying causes and viable solutions. By analyzing the study results, we aimed to provide insights for Copilot users, the Copilot team, and researchers. Besides, we collected data from three popular software development platforms and forums, i.e., GitHub Issues, GitHub Discussions, and SO, to ensure the comprehensiveness of our dataset.

\section{Conclusions}
\label{sec:conclusions}
In this study, we focused on the problems users may encounter when using GitHub Copilot, as well as their underlying causes and potential solutions. After identifying the RQs, we collected data from GitHub Issues, GitHub Discussions, and Stack Overflow. After manual screening, we obtained 473 GitHub issues, 706 GitHub discussions, and 142 SO posts related to Copilot and got a total of 1353 problems, 391 causes, and 497 solutions based on our data extraction criteria. The results indicate that \textit{Operation Issue} and \textit{Compatibility Issue} are the most common problems faced by users. \textit{Copilot Internal Error}, \textit{Network Connection Error}, and \textit{Editor/IDE Compatibility Issue} are identified as the most common causes of these problems. \textit{Bug Fixed by Copilot}, \textit{Modify Configuration/Setting} and \textit{Use Suitable Version} are the predominant solution. %Our findings suggest that the decision to keep Copilot closed-source has both positive and negative implications for product development and user experience. In addition to improving the quality of code suggestions, the Copilot team needs to allocate more resources towards addressing compatibility and configuration issues across different development environments. Furthermore, prioritizing user feedback could facilitate the exploration of more effective ways for software development to integrate with AI-powered code generation tools.
% Our findings suggest that Copilot should enhance compatibility across various IDEs and editors, simplify the configuration, improve the quality of generated code, and address concerns related to intellectual property and copyright. Additionally, users require more customization options to tailor Copilot's behavior and have more control over the content generated by Copilot. In light of the additional time required for code suggestion verification when utilizing Copilot, the integration of a code explanation feature becomes imperative to enhance its overall utility and effectiveness in practical development scenarios.
For the Copilot users, our results suggest that they should carefully review its code suggestions and seek inspiration from them. IDEs or code editors that are officially supported by Copilot can lead to a better user experience. For the Copilot team, it is essential to enhance the compatibility and provide support for a broader range of IDEs and code editors, simplify the configuration process, diversify the code suggestions while improving their quality, enhance Copilot performance for niche or domain-specific programming languages, and address concerns related to intellectual property and copyright. Additionally, users are asking for more customization options to tailor Copilot's behavior and more control over the code content generated by Copilot. For researchers, we found that additional time is required for code suggestion verification when utilizing Copilot, thus making code explanation feature especially valuable. Programming tasks in different application domains and the purpose for which users employ Copilot should be taken into consideration when assessing user satisfaction with Copilot. Consistently monitoring the feedback in \textit{Feature Request}, particularly after introducing new features, can guide future improvements and identify specific areas for enhancing Copilot.

% In the next step, we plan to conduct an industrial survey with code testing experiments to assess the real-world usage of Copilot by users, as well as its performance in terms of security, maintainability, and other aspects. Considering the emergence of LLMs is likely to drive a significant proliferation of AI code assistant tools, a comparison between Copilot and other LLM-based tools holds valuable insights.

For future work, we plan to continuously collect new data from the three data sources to analyze the latest usage problems that occur during the iterations of Copilot, along with their causes and solutions. With the introduction of GitHub Copilot Trust Center~\citep{copilottrustcenter}, we plan to investigate whether the concerns of users for \textit{Copyright and Policy Issue} will be addressed. Besides, we will employ an industrial survey to gather more information, such as the programming experiences of Copilot users and the development contexts when coding with Copilot, to further investigate how these factors impact the categories of Copilot-related problems encountered by users. Additionally, we aim to understand the impact of Copilot on various stages of software development life cycle by conducting an interview with developers who use Copilot in industrial software development. Besides, considering the emergence of LLMs is likely to drive a significant proliferation of AI code assistant tools, a comparison between Copilot and other popular AI code assistant tools (e.g., Tabnine \citep{tabnine}, Amazon Q Developer \citep{amazonqdeveloper}) holds valuable insights.

\section*{Data availability}
We have shared the link to our dataset in the reference \citep{dataset}.

%\section*{conflict of interest}
%You may be asked to provide a conflict of interest statement during the submission process. Please check the journal's author guidelines for details on what to include in this section. Please ensure you liaise with all co-authors to confirm agreement with the final statement.

\section*{Acknowledgments}
This work is supported by the National Natural Science Foundation of China under Grant Nos. 62172311 and 62176099, the Natural Science Foundation of Hubei Province of China under Grant No. 2021CFB577, and the Knowledge Innovation Program of Wuhan-Shuguang Project under Grant No. 2022010801020280.

\printcredits

% \appendix
% \section{Appendix: Abbreviations used in the study}\label{appendixA}
% \begin{tabular}{p{0.15\columnwidth} p{0.7\columnwidth}}
%     %AveC&Average Value of the Corresponding Change Complexity Metric of Bug-Fixing Commits (or Communication Complexity Metrics) for Bugs with Priority Changes    \\
%     %AveN&Average Value of the Corresponding Change Complexity Metric of Bug-Fixing Commits (or Communication Complexity Metrics) for Bugs without Priority Changes    \\
%     BugPC&Bug with Priority Changes    \\
%     GQM&Goal-Question-Metric    \\
%     LOCM&Lines of Code Modified    \\
%     $N_a$&Average Number of Priorities Allocated    \\
%     $N_c$&Average Number of Bugs Involved by Each Core Participant    \\
%     NOC&Number of Comments    \\
%     NOCR&Mumber of Commenters    \\
%     NOFM&Number of Files (for Java) Modified    \\
%     NOPM&Number of Packages (for Java) Modified    \\
%     $N_r$&Average Number of Bugs Reported    \\
%     OR&Original Priority    \\
%     OSS&Open Source Software    \\
%     PA&Participants whose Allocating Bug Priorities Are Likely to be Modified    \\
%     PC&Priority Change    \\
%     PM&Participants who Are Likely to Modify Bug Priorities    \\
%     PR&Participant whose Reporting Bug Priorities Are Likely to be Modified    \\
%     RQ&Research Question  \\
%     TLC&Total Length of All Comments
% \end{tabular}

\bibliographystyle{cas-model2-names}

\bibliography{references}
\balance
\end{sloppypar}
\end{document}